\documentclass[longauth]{aaEC}

\usepackage{graphicx}
\usepackage{natbib}
\usepackage{scalerel}

\usepackage[table]{xcolor}

\usepackage{txfonts}
\usepackage[pdfencoding=auto,psdextra]{hyperref}
\hypersetup{
    colorlinks=true,
    linkcolor=blue,
    filecolor=magenta,      
    urlcolor=blue,
    citecolor=blue
}
\usepackage{orcidlink}

\makeatletter
\renewcommand*\aa@pageof{, page \thepage{} of \pageref*{LastPage}}
\makeatother
\usepackage[utf8]{inputenc}

\usepackage[switch, modulo]{lineno}

\usepackage{euclid}

\begin{document}
%
%


\title{Euclid Quick Data Release (Q1)}
\subtitle{First visual morphology catalogue}    

\newcommand{\orcid}[1]{\href{https://orcid.org/#1}{\orcidlink{#1}}}   
\author{Euclid Collaboration: M.~Walmsley\orcid{0000-0002-6408-4181}\thanks{\email{m.walmsley@utoronto.ca}}\inst{\ref{aff1},\ref{aff2}}
\and M.~Huertas-Company\orcid{0000-0002-1416-8483}\inst{\ref{aff3},\ref{aff4},\ref{aff5},\ref{aff6}}
\and L.~Quilley\orcid{0009-0008-8375-8605}\inst{\ref{aff7}}
\and K.~L.~Masters\orcid{0000-0003-0846-9578}\inst{\ref{aff8}}
\and S.~Kruk\orcid{0000-0001-8010-8879}\inst{\ref{aff9}}
\and K.~A.~Remmelgas\inst{\ref{aff9}}
\and J.~J.~Popp\orcid{0000-0002-3724-1727}\inst{\ref{aff10}}
\and E.~Romelli\orcid{0000-0003-3069-9222}\inst{\ref{aff11}}
\and D.~O'Ryan\orcid{0000-0003-1217-4617}\inst{\ref{aff12}}
\and H.~J.~Dickinson\orcid{0000-0003-0475-008X}\inst{\ref{aff10}}
\and C.~J.~Lintott\orcid{0000-0001-5578-359X}\inst{\ref{aff13}}
\and S.~Serjeant\orcid{0000-0002-0517-7943}\inst{\ref{aff10}}
\and R.~J.~Smethurst\orcid{0000-0001-6417-7196}\inst{\ref{aff13}}
\and B.~Simmons\orcid{0000-0001-5882-3323}\inst{\ref{aff14}}
\and J.~Shingirai~Makechemu\orcid{0009-0009-6545-8710}\inst{\ref{aff14}}
\and I.~L.~Garland\orcid{0000-0002-3887-6433}\inst{\ref{aff15}}
\and H.~Roberts\orcid{0000-0003-0046-9848}\inst{\ref{aff16}}
\and K.~Mantha\orcid{0000-0002-6016-300X}\inst{\ref{aff16}}
\and L.~F.~Fortson\orcid{0000-0002-1067-8558}\inst{\ref{aff16}}
\and T.~G\'eron\orcid{0000-0002-6851-9613}\inst{\ref{aff1}}
\and W.~Keel\orcid{0000-0002-6131-9539}\inst{\ref{aff17}}
\and E.~M.~Baeten\inst{\ref{aff18}}
\and C.~Macmillan\inst{\ref{aff13}}
\and J.~Bovy\orcid{0000-0001-6855-442X}\inst{\ref{aff1}}
\and S.~Casas\orcid{0000-0002-4751-5138}\inst{\ref{aff19}}
\and C.~De~Leo\inst{\ref{aff20}}
\and H.~Dom\'inguez~S\'anchez\orcid{0000-0002-9013-1316}\inst{\ref{aff21}}
\and J.~Katona\orcid{0009-0001-5371-8935}\inst{\ref{aff22},\ref{aff23}}
\and A.~Kov\'acs\orcid{0000-0002-5825-579X}\inst{\ref{aff23},\ref{aff24}}
\and N.~Aghanim\orcid{0000-0002-6688-8992}\inst{\ref{aff25}}
\and B.~Altieri\orcid{0000-0003-3936-0284}\inst{\ref{aff9}}
\and A.~Amara\inst{\ref{aff26}}
\and S.~Andreon\orcid{0000-0002-2041-8784}\inst{\ref{aff27}}
\and N.~Auricchio\orcid{0000-0003-4444-8651}\inst{\ref{aff28}}
\and H.~Aussel\orcid{0000-0002-1371-5705}\inst{\ref{aff29}}
\and C.~Baccigalupi\orcid{0000-0002-8211-1630}\inst{\ref{aff30},\ref{aff11},\ref{aff31},\ref{aff32}}
\and M.~Baldi\orcid{0000-0003-4145-1943}\inst{\ref{aff33},\ref{aff28},\ref{aff34}}
\and A.~Balestra\orcid{0000-0002-6967-261X}\inst{\ref{aff35}}
\and S.~Bardelli\orcid{0000-0002-8900-0298}\inst{\ref{aff28}}
\and A.~Basset\inst{\ref{aff36}}
\and P.~Battaglia\orcid{0000-0002-7337-5909}\inst{\ref{aff28}}
\and R.~Bender\orcid{0000-0001-7179-0626}\inst{\ref{aff37},\ref{aff38}}
\and A.~Biviano\orcid{0000-0002-0857-0732}\inst{\ref{aff11},\ref{aff30}}
\and A.~Bonchi\orcid{0000-0002-2667-5482}\inst{\ref{aff39}}
\and E.~Branchini\orcid{0000-0002-0808-6908}\inst{\ref{aff40},\ref{aff41},\ref{aff27}}
\and M.~Brescia\orcid{0000-0001-9506-5680}\inst{\ref{aff42},\ref{aff43}}
\and J.~Brinchmann\orcid{0000-0003-4359-8797}\inst{\ref{aff44},\ref{aff45}}
\and S.~Camera\orcid{0000-0003-3399-3574}\inst{\ref{aff46},\ref{aff47},\ref{aff48}}
\and G.~Ca\~nas-Herrera\orcid{0000-0003-2796-2149}\inst{\ref{aff49},\ref{aff50},\ref{aff51}}
\and V.~Capobianco\orcid{0000-0002-3309-7692}\inst{\ref{aff48}}
\and C.~Carbone\orcid{0000-0003-0125-3563}\inst{\ref{aff52}}
\and J.~Carretero\orcid{0000-0002-3130-0204}\inst{\ref{aff53},\ref{aff54}}
\and F.~J.~Castander\orcid{0000-0001-7316-4573}\inst{\ref{aff55},\ref{aff56}}
\and M.~Castellano\orcid{0000-0001-9875-8263}\inst{\ref{aff57}}
\and G.~Castignani\orcid{0000-0001-6831-0687}\inst{\ref{aff28}}
\and S.~Cavuoti\orcid{0000-0002-3787-4196}\inst{\ref{aff43},\ref{aff58}}
\and K.~C.~Chambers\orcid{0000-0001-6965-7789}\inst{\ref{aff59}}
\and A.~Cimatti\inst{\ref{aff60}}
\and C.~Colodro-Conde\inst{\ref{aff3}}
\and G.~Congedo\orcid{0000-0003-2508-0046}\inst{\ref{aff61}}
\and C.~J.~Conselice\orcid{0000-0003-1949-7638}\inst{\ref{aff2}}
\and L.~Conversi\orcid{0000-0002-6710-8476}\inst{\ref{aff62},\ref{aff9}}
\and Y.~Copin\orcid{0000-0002-5317-7518}\inst{\ref{aff63}}
\and F.~Courbin\orcid{0000-0003-0758-6510}\inst{\ref{aff64},\ref{aff65}}
\and H.~M.~Courtois\orcid{0000-0003-0509-1776}\inst{\ref{aff66}}
\and M.~Cropper\orcid{0000-0003-4571-9468}\inst{\ref{aff67}}
\and A.~Da~Silva\orcid{0000-0002-6385-1609}\inst{\ref{aff68},\ref{aff69}}
\and H.~Degaudenzi\orcid{0000-0002-5887-6799}\inst{\ref{aff70}}
\and G.~De~Lucia\orcid{0000-0002-6220-9104}\inst{\ref{aff11}}
\and A.~M.~Di~Giorgio\orcid{0000-0002-4767-2360}\inst{\ref{aff71}}
\and C.~Dolding\orcid{0009-0003-7199-6108}\inst{\ref{aff67}}
\and H.~Dole\orcid{0000-0002-9767-3839}\inst{\ref{aff25}}
\and F.~Dubath\orcid{0000-0002-6533-2810}\inst{\ref{aff70}}
\and C.~A.~J.~Duncan\orcid{0009-0003-3573-0791}\inst{\ref{aff2}}
\and X.~Dupac\inst{\ref{aff9}}
\and S.~Dusini\orcid{0000-0002-1128-0664}\inst{\ref{aff72}}
\and A.~Ealet\orcid{0000-0003-3070-014X}\inst{\ref{aff63}}
\and S.~Escoffier\orcid{0000-0002-2847-7498}\inst{\ref{aff73}}
\and M.~Fabricius\orcid{0000-0002-7025-6058}\inst{\ref{aff37},\ref{aff38}}
\and M.~Farina\orcid{0000-0002-3089-7846}\inst{\ref{aff71}}
\and R.~Farinelli\inst{\ref{aff28}}
\and F.~Faustini\orcid{0000-0001-6274-5145}\inst{\ref{aff39},\ref{aff57}}
\and F.~Finelli\orcid{0000-0002-6694-3269}\inst{\ref{aff28},\ref{aff74}}
\and P.~Fosalba\orcid{0000-0002-1510-5214}\inst{\ref{aff56},\ref{aff55}}
\and S.~Fotopoulou\orcid{0000-0002-9686-254X}\inst{\ref{aff75}}
\and M.~Frailis\orcid{0000-0002-7400-2135}\inst{\ref{aff11}}
\and E.~Franceschi\orcid{0000-0002-0585-6591}\inst{\ref{aff28}}
\and S.~Galeotta\orcid{0000-0002-3748-5115}\inst{\ref{aff11}}
\and K.~George\orcid{0000-0002-1734-8455}\inst{\ref{aff38}}
\and B.~Gillis\orcid{0000-0002-4478-1270}\inst{\ref{aff61}}
\and C.~Giocoli\orcid{0000-0002-9590-7961}\inst{\ref{aff28},\ref{aff34}}
\and P.~G\'omez-Alvarez\orcid{0000-0002-8594-5358}\inst{\ref{aff76},\ref{aff9}}
\and J.~Gracia-Carpio\inst{\ref{aff37}}
\and B.~R.~Granett\orcid{0000-0003-2694-9284}\inst{\ref{aff27}}
\and A.~Grazian\orcid{0000-0002-5688-0663}\inst{\ref{aff35}}
\and F.~Grupp\inst{\ref{aff37},\ref{aff38}}
\and S.~Gwyn\orcid{0000-0001-8221-8406}\inst{\ref{aff77}}
\and S.~V.~H.~Haugan\orcid{0000-0001-9648-7260}\inst{\ref{aff78}}
\and H.~Hoekstra\orcid{0000-0002-0641-3231}\inst{\ref{aff51}}
\and W.~Holmes\inst{\ref{aff79}}
\and I.~M.~Hook\orcid{0000-0002-2960-978X}\inst{\ref{aff14}}
\and F.~Hormuth\inst{\ref{aff80}}
\and A.~Hornstrup\orcid{0000-0002-3363-0936}\inst{\ref{aff81},\ref{aff82}}
\and P.~Hudelot\inst{\ref{aff83}}
\and K.~Jahnke\orcid{0000-0003-3804-2137}\inst{\ref{aff84}}
\and M.~Jhabvala\inst{\ref{aff85}}
\and B.~Joachimi\orcid{0000-0001-7494-1303}\inst{\ref{aff86}}
\and E.~Keih\"anen\orcid{0000-0003-1804-7715}\inst{\ref{aff87}}
\and S.~Kermiche\orcid{0000-0002-0302-5735}\inst{\ref{aff73}}
\and A.~Kiessling\orcid{0000-0002-2590-1273}\inst{\ref{aff79}}
\and R.~Kohley\inst{\ref{aff9}}
\and B.~Kubik\orcid{0009-0006-5823-4880}\inst{\ref{aff63}}
\and K.~Kuijken\orcid{0000-0002-3827-0175}\inst{\ref{aff51}}
\and M.~K\"ummel\orcid{0000-0003-2791-2117}\inst{\ref{aff38}}
\and M.~Kunz\orcid{0000-0002-3052-7394}\inst{\ref{aff88}}
\and H.~Kurki-Suonio\orcid{0000-0002-4618-3063}\inst{\ref{aff89},\ref{aff90}}
\and O.~Lahav\orcid{0000-0002-1134-9035}\inst{\ref{aff86}}
\and Q.~Le~Boulc'h\inst{\ref{aff91}}
\and A.~M.~C.~Le~Brun\orcid{0000-0002-0936-4594}\inst{\ref{aff92}}
\and D.~Le~Mignant\orcid{0000-0002-5339-5515}\inst{\ref{aff93}}
\and P.~Liebing\inst{\ref{aff67}}
\and S.~Ligori\orcid{0000-0003-4172-4606}\inst{\ref{aff48}}
\and P.~B.~Lilje\orcid{0000-0003-4324-7794}\inst{\ref{aff78}}
\and V.~Lindholm\orcid{0000-0003-2317-5471}\inst{\ref{aff89},\ref{aff90}}
\and I.~Lloro\orcid{0000-0001-5966-1434}\inst{\ref{aff94}}
\and G.~Mainetti\orcid{0000-0003-2384-2377}\inst{\ref{aff91}}
\and D.~Maino\inst{\ref{aff95},\ref{aff52},\ref{aff96}}
\and E.~Maiorano\orcid{0000-0003-2593-4355}\inst{\ref{aff28}}
\and O.~Mansutti\orcid{0000-0001-5758-4658}\inst{\ref{aff11}}
\and S.~Marcin\inst{\ref{aff97}}
\and O.~Marggraf\orcid{0000-0001-7242-3852}\inst{\ref{aff98}}
\and M.~Martinelli\orcid{0000-0002-6943-7732}\inst{\ref{aff57},\ref{aff99}}
\and N.~Martinet\orcid{0000-0003-2786-7790}\inst{\ref{aff93}}
\and F.~Marulli\orcid{0000-0002-8850-0303}\inst{\ref{aff100},\ref{aff28},\ref{aff34}}
\and R.~Massey\orcid{0000-0002-6085-3780}\inst{\ref{aff101}}
\and S.~Maurogordato\inst{\ref{aff102}}
\and H.~J.~McCracken\orcid{0000-0002-9489-7765}\inst{\ref{aff83}}
\and E.~Medinaceli\orcid{0000-0002-4040-7783}\inst{\ref{aff28}}
\and S.~Mei\orcid{0000-0002-2849-559X}\inst{\ref{aff103},\ref{aff104}}
\and M.~Melchior\inst{\ref{aff105}}
\and Y.~Mellier\inst{\ref{aff106},\ref{aff83}}
\and M.~Meneghetti\orcid{0000-0003-1225-7084}\inst{\ref{aff28},\ref{aff34}}
\and E.~Merlin\orcid{0000-0001-6870-8900}\inst{\ref{aff57}}
\and G.~Meylan\inst{\ref{aff107}}
\and A.~Mora\orcid{0000-0002-1922-8529}\inst{\ref{aff108}}
\and M.~Moresco\orcid{0000-0002-7616-7136}\inst{\ref{aff100},\ref{aff28}}
\and L.~Moscardini\orcid{0000-0002-3473-6716}\inst{\ref{aff100},\ref{aff28},\ref{aff34}}
\and R.~Nakajima\orcid{0009-0009-1213-7040}\inst{\ref{aff98}}
\and C.~Neissner\orcid{0000-0001-8524-4968}\inst{\ref{aff109},\ref{aff54}}
\and R.~C.~Nichol\orcid{0000-0003-0939-6518}\inst{\ref{aff26}}
\and S.-M.~Niemi\inst{\ref{aff49}}
\and J.~W.~Nightingale\orcid{0000-0002-8987-7401}\inst{\ref{aff110}}
\and C.~Padilla\orcid{0000-0001-7951-0166}\inst{\ref{aff109}}
\and S.~Paltani\orcid{0000-0002-8108-9179}\inst{\ref{aff70}}
\and F.~Pasian\orcid{0000-0002-4869-3227}\inst{\ref{aff11}}
\and K.~Pedersen\inst{\ref{aff111}}
\and W.~J.~Percival\orcid{0000-0002-0644-5727}\inst{\ref{aff112},\ref{aff113},\ref{aff114}}
\and V.~Pettorino\inst{\ref{aff49}}
\and S.~Pires\orcid{0000-0002-0249-2104}\inst{\ref{aff29}}
\and G.~Polenta\orcid{0000-0003-4067-9196}\inst{\ref{aff39}}
\and M.~Poncet\inst{\ref{aff36}}
\and L.~A.~Popa\inst{\ref{aff115}}
\and L.~Pozzetti\orcid{0000-0001-7085-0412}\inst{\ref{aff28}}
\and F.~Raison\orcid{0000-0002-7819-6918}\inst{\ref{aff37}}
\and R.~Rebolo\orcid{0000-0003-3767-7085}\inst{\ref{aff3},\ref{aff116},\ref{aff117}}
\and A.~Renzi\orcid{0000-0001-9856-1970}\inst{\ref{aff118},\ref{aff72}}
\and J.~Rhodes\orcid{0000-0002-4485-8549}\inst{\ref{aff79}}
\and G.~Riccio\inst{\ref{aff43}}
\and M.~Roncarelli\orcid{0000-0001-9587-7822}\inst{\ref{aff28}}
\and B.~Rusholme\orcid{0000-0001-7648-4142}\inst{\ref{aff119}}
\and R.~Saglia\orcid{0000-0003-0378-7032}\inst{\ref{aff38},\ref{aff37}}
\and Z.~Sakr\orcid{0000-0002-4823-3757}\inst{\ref{aff120},\ref{aff121},\ref{aff122}}
\and A.~G.~S\'anchez\orcid{0000-0003-1198-831X}\inst{\ref{aff37}}
\and D.~Sapone\orcid{0000-0001-7089-4503}\inst{\ref{aff123}}
\and B.~Sartoris\orcid{0000-0003-1337-5269}\inst{\ref{aff38},\ref{aff11}}
\and J.~A.~Schewtschenko\orcid{0000-0002-4913-6393}\inst{\ref{aff61}}
\and P.~Schneider\orcid{0000-0001-8561-2679}\inst{\ref{aff98}}
\and T.~Schrabback\orcid{0000-0002-6987-7834}\inst{\ref{aff124}}
\and M.~Scodeggio\inst{\ref{aff52}}
\and A.~Secroun\orcid{0000-0003-0505-3710}\inst{\ref{aff73}}
\and G.~Seidel\orcid{0000-0003-2907-353X}\inst{\ref{aff84}}
\and M.~Seiffert\orcid{0000-0002-7536-9393}\inst{\ref{aff79}}
\and S.~Serrano\orcid{0000-0002-0211-2861}\inst{\ref{aff56},\ref{aff125},\ref{aff55}}
\and P.~Simon\inst{\ref{aff98}}
\and C.~Sirignano\orcid{0000-0002-0995-7146}\inst{\ref{aff118},\ref{aff72}}
\and G.~Sirri\orcid{0000-0003-2626-2853}\inst{\ref{aff34}}
\and L.~Stanco\orcid{0000-0002-9706-5104}\inst{\ref{aff72}}
\and J.~Steinwagner\orcid{0000-0001-7443-1047}\inst{\ref{aff37}}
\and P.~Tallada-Cresp\'{i}\orcid{0000-0002-1336-8328}\inst{\ref{aff53},\ref{aff54}}
\and D.~Tavagnacco\orcid{0000-0001-7475-9894}\inst{\ref{aff11}}
\and A.~N.~Taylor\inst{\ref{aff61}}
\and H.~I.~Teplitz\orcid{0000-0002-7064-5424}\inst{\ref{aff126}}
\and I.~Tereno\inst{\ref{aff68},\ref{aff127}}
\and N.~Tessore\orcid{0000-0002-9696-7931}\inst{\ref{aff86}}
\and S.~Toft\orcid{0000-0003-3631-7176}\inst{\ref{aff128},\ref{aff129}}
\and R.~Toledo-Moreo\orcid{0000-0002-2997-4859}\inst{\ref{aff130}}
\and F.~Torradeflot\orcid{0000-0003-1160-1517}\inst{\ref{aff54},\ref{aff53}}
\and I.~Tutusaus\orcid{0000-0002-3199-0399}\inst{\ref{aff121}}
\and E.~A.~Valentijn\inst{\ref{aff131}}
\and L.~Valenziano\orcid{0000-0002-1170-0104}\inst{\ref{aff28},\ref{aff74}}
\and J.~Valiviita\orcid{0000-0001-6225-3693}\inst{\ref{aff89},\ref{aff90}}
\and T.~Vassallo\orcid{0000-0001-6512-6358}\inst{\ref{aff38},\ref{aff11}}
\and G.~Verdoes~Kleijn\orcid{0000-0001-5803-2580}\inst{\ref{aff131}}
\and A.~Veropalumbo\orcid{0000-0003-2387-1194}\inst{\ref{aff27},\ref{aff41},\ref{aff40}}
\and Y.~Wang\orcid{0000-0002-4749-2984}\inst{\ref{aff126}}
\and J.~Weller\orcid{0000-0002-8282-2010}\inst{\ref{aff38},\ref{aff37}}
\and A.~Zacchei\orcid{0000-0003-0396-1192}\inst{\ref{aff11},\ref{aff30}}
\and G.~Zamorani\orcid{0000-0002-2318-301X}\inst{\ref{aff28}}
\and F.~M.~Zerbi\inst{\ref{aff27}}
\and I.~A.~Zinchenko\orcid{0000-0002-2944-2449}\inst{\ref{aff38}}
\and E.~Zucca\orcid{0000-0002-5845-8132}\inst{\ref{aff28}}
\and V.~Allevato\orcid{0000-0001-7232-5152}\inst{\ref{aff43}}
\and M.~Ballardini\orcid{0000-0003-4481-3559}\inst{\ref{aff132},\ref{aff133},\ref{aff28}}
\and M.~Bolzonella\orcid{0000-0003-3278-4607}\inst{\ref{aff28}}
\and E.~Bozzo\orcid{0000-0002-8201-1525}\inst{\ref{aff70}}
\and C.~Burigana\orcid{0000-0002-3005-5796}\inst{\ref{aff134},\ref{aff74}}
\and R.~Cabanac\orcid{0000-0001-6679-2600}\inst{\ref{aff121}}
\and A.~Cappi\inst{\ref{aff28},\ref{aff102}}
\and D.~Di~Ferdinando\inst{\ref{aff34}}
\and J.~A.~Escartin~Vigo\inst{\ref{aff37}}
\and L.~Gabarra\orcid{0000-0002-8486-8856}\inst{\ref{aff13}}
\and J.~Mart\'{i}n-Fleitas\orcid{0000-0002-8594-569X}\inst{\ref{aff108}}
\and S.~Matthew\orcid{0000-0001-8448-1697}\inst{\ref{aff61}}
\and N.~Mauri\orcid{0000-0001-8196-1548}\inst{\ref{aff60},\ref{aff34}}
\and R.~B.~Metcalf\orcid{0000-0003-3167-2574}\inst{\ref{aff100},\ref{aff28}}
\and A.~Pezzotta\orcid{0000-0003-0726-2268}\inst{\ref{aff135},\ref{aff37}}
\and M.~P\"ontinen\orcid{0000-0001-5442-2530}\inst{\ref{aff89}}
\and C.~Porciani\orcid{0000-0002-7797-2508}\inst{\ref{aff98}}
\and I.~Risso\orcid{0000-0003-2525-7761}\inst{\ref{aff136}}
\and V.~Scottez\inst{\ref{aff106},\ref{aff137}}
\and M.~Sereno\orcid{0000-0003-0302-0325}\inst{\ref{aff28},\ref{aff34}}
\and M.~Tenti\orcid{0000-0002-4254-5901}\inst{\ref{aff34}}
\and M.~Viel\orcid{0000-0002-2642-5707}\inst{\ref{aff30},\ref{aff11},\ref{aff32},\ref{aff31},\ref{aff138}}
\and M.~Wiesmann\orcid{0009-0000-8199-5860}\inst{\ref{aff78}}
\and Y.~Akrami\orcid{0000-0002-2407-7956}\inst{\ref{aff139},\ref{aff140}}
\and I.~T.~Andika\orcid{0000-0001-6102-9526}\inst{\ref{aff141},\ref{aff142}}
\and S.~Anselmi\orcid{0000-0002-3579-9583}\inst{\ref{aff72},\ref{aff118},\ref{aff143}}
\and M.~Archidiacono\orcid{0000-0003-4952-9012}\inst{\ref{aff95},\ref{aff96}}
\and F.~Atrio-Barandela\orcid{0000-0002-2130-2513}\inst{\ref{aff144}}
\and C.~Benoist\inst{\ref{aff102}}
\and K.~Benson\inst{\ref{aff67}}
\and D.~Bertacca\orcid{0000-0002-2490-7139}\inst{\ref{aff118},\ref{aff35},\ref{aff72}}
\and M.~Bethermin\orcid{0000-0002-3915-2015}\inst{\ref{aff145}}
\and L.~Bisigello\orcid{0000-0003-0492-4924}\inst{\ref{aff35}}
\and A.~Blanchard\orcid{0000-0001-8555-9003}\inst{\ref{aff121}}
\and L.~Blot\orcid{0000-0002-9622-7167}\inst{\ref{aff146},\ref{aff143}}
\and H.~B\"ohringer\orcid{0000-0001-8241-4204}\inst{\ref{aff37},\ref{aff147},\ref{aff148}}
\and M.~L.~Brown\orcid{0000-0002-0370-8077}\inst{\ref{aff2}}
\and S.~Bruton\orcid{0000-0002-6503-5218}\inst{\ref{aff149}}
\and F.~Buitrago\orcid{0000-0002-2861-9812}\inst{\ref{aff150},\ref{aff127}}
\and A.~Calabro\orcid{0000-0003-2536-1614}\inst{\ref{aff57}}
\and B.~Camacho~Quevedo\orcid{0000-0002-8789-4232}\inst{\ref{aff56},\ref{aff55}}
\and F.~Caro\inst{\ref{aff57}}
\and C.~S.~Carvalho\inst{\ref{aff127}}
\and T.~Castro\orcid{0000-0002-6292-3228}\inst{\ref{aff11},\ref{aff31},\ref{aff30},\ref{aff138}}
\and F.~Cogato\orcid{0000-0003-4632-6113}\inst{\ref{aff100},\ref{aff28}}
\and A.~R.~Cooray\orcid{0000-0002-3892-0190}\inst{\ref{aff151}}
\and O.~Cucciati\orcid{0000-0002-9336-7551}\inst{\ref{aff28}}
\and S.~Davini\orcid{0000-0003-3269-1718}\inst{\ref{aff41}}
\and F.~De~Paolis\orcid{0000-0001-6460-7563}\inst{\ref{aff152},\ref{aff153},\ref{aff154}}
\and G.~Desprez\orcid{0000-0001-8325-1742}\inst{\ref{aff131}}
\and A.~D\'iaz-S\'anchez\orcid{0000-0003-0748-4768}\inst{\ref{aff155}}
\and J.~J.~Diaz\inst{\ref{aff4},\ref{aff3}}
\and S.~Di~Domizio\orcid{0000-0003-2863-5895}\inst{\ref{aff40},\ref{aff41}}
\and J.~M.~Diego\orcid{0000-0001-9065-3926}\inst{\ref{aff21}}
\and P.-A.~Duc\orcid{0000-0003-3343-6284}\inst{\ref{aff145}}
\and A.~Enia\orcid{0000-0002-0200-2857}\inst{\ref{aff33},\ref{aff28}}
\and Y.~Fang\inst{\ref{aff38}}
\and A.~G.~Ferrari\orcid{0009-0005-5266-4110}\inst{\ref{aff34}}
\and A.~Finoguenov\orcid{0000-0002-4606-5403}\inst{\ref{aff89}}
\and A.~Fontana\orcid{0000-0003-3820-2823}\inst{\ref{aff57}}
\and A.~Franco\orcid{0000-0002-4761-366X}\inst{\ref{aff153},\ref{aff152},\ref{aff154}}
\and K.~Ganga\orcid{0000-0001-8159-8208}\inst{\ref{aff103}}
\and J.~Garc\'ia-Bellido\orcid{0000-0002-9370-8360}\inst{\ref{aff139}}
\and T.~Gasparetto\orcid{0000-0002-7913-4866}\inst{\ref{aff11}}
\and V.~Gautard\inst{\ref{aff156}}
\and E.~Gaztanaga\orcid{0000-0001-9632-0815}\inst{\ref{aff55},\ref{aff56},\ref{aff157}}
\and F.~Giacomini\orcid{0000-0002-3129-2814}\inst{\ref{aff34}}
\and G.~Gozaliasl\orcid{0000-0002-0236-919X}\inst{\ref{aff158},\ref{aff89}}
\and M.~Guidi\orcid{0000-0001-9408-1101}\inst{\ref{aff33},\ref{aff28}}
\and C.~M.~Gutierrez\orcid{0000-0001-7854-783X}\inst{\ref{aff159}}
\and A.~Hall\orcid{0000-0002-3139-8651}\inst{\ref{aff61}}
\and W.~G.~Hartley\inst{\ref{aff70}}
\and S.~Hemmati\orcid{0000-0003-2226-5395}\inst{\ref{aff119}}
\and C.~Hern\'andez-Monteagudo\orcid{0000-0001-5471-9166}\inst{\ref{aff117},\ref{aff3}}
\and H.~Hildebrandt\orcid{0000-0002-9814-3338}\inst{\ref{aff160}}
\and J.~Hjorth\orcid{0000-0002-4571-2306}\inst{\ref{aff111}}
\and J.~J.~E.~Kajava\orcid{0000-0002-3010-8333}\inst{\ref{aff161},\ref{aff162}}
\and Y.~Kang\orcid{0009-0000-8588-7250}\inst{\ref{aff70}}
\and V.~Kansal\orcid{0000-0002-4008-6078}\inst{\ref{aff163},\ref{aff164}}
\and D.~Karagiannis\orcid{0000-0002-4927-0816}\inst{\ref{aff132},\ref{aff165}}
\and K.~Kiiveri\inst{\ref{aff87}}
\and C.~C.~Kirkpatrick\inst{\ref{aff87}}
\and J.~Le~Graet\orcid{0000-0001-6523-7971}\inst{\ref{aff73}}
\and L.~Legrand\orcid{0000-0003-0610-5252}\inst{\ref{aff166},\ref{aff167}}
\and M.~Lembo\orcid{0000-0002-5271-5070}\inst{\ref{aff132},\ref{aff133}}
\and F.~Lepori\orcid{0009-0000-5061-7138}\inst{\ref{aff168}}
\and G.~Leroy\orcid{0009-0004-2523-4425}\inst{\ref{aff169},\ref{aff101}}
\and G.~F.~Lesci\orcid{0000-0002-4607-2830}\inst{\ref{aff100},\ref{aff28}}
\and J.~Lesgourgues\orcid{0000-0001-7627-353X}\inst{\ref{aff19}}
\and L.~Leuzzi\orcid{0009-0006-4479-7017}\inst{\ref{aff100},\ref{aff28}}
\and T.~I.~Liaudat\orcid{0000-0002-9104-314X}\inst{\ref{aff170}}
\and A.~Loureiro\orcid{0000-0002-4371-0876}\inst{\ref{aff171},\ref{aff172}}
\and J.~Macias-Perez\orcid{0000-0002-5385-2763}\inst{\ref{aff173}}
\and G.~Maggio\orcid{0000-0003-4020-4836}\inst{\ref{aff11}}
\and M.~Magliocchetti\orcid{0000-0001-9158-4838}\inst{\ref{aff71}}
\and F.~Mannucci\orcid{0000-0002-4803-2381}\inst{\ref{aff174}}
\and R.~Maoli\orcid{0000-0002-6065-3025}\inst{\ref{aff20},\ref{aff57}}
\and C.~J.~A.~P.~Martins\orcid{0000-0002-4886-9261}\inst{\ref{aff175},\ref{aff44}}
\and L.~Maurin\orcid{0000-0002-8406-0857}\inst{\ref{aff25}}
\and M.~Miluzio\inst{\ref{aff9},\ref{aff176}}
\and P.~Monaco\orcid{0000-0003-2083-7564}\inst{\ref{aff177},\ref{aff11},\ref{aff31},\ref{aff30}}
\and C.~Moretti\orcid{0000-0003-3314-8936}\inst{\ref{aff32},\ref{aff138},\ref{aff11},\ref{aff30},\ref{aff31}}
\and G.~Morgante\inst{\ref{aff28}}
\and C.~Murray\inst{\ref{aff103}}
\and S.~Nadathur\orcid{0000-0001-9070-3102}\inst{\ref{aff157}}
\and K.~Naidoo\orcid{0000-0002-9182-1802}\inst{\ref{aff157}}
\and A.~Navarro-Alsina\orcid{0000-0002-3173-2592}\inst{\ref{aff98}}
\and S.~Nesseris\orcid{0000-0002-0567-0324}\inst{\ref{aff139}}
\and F.~Passalacqua\orcid{0000-0002-8606-4093}\inst{\ref{aff118},\ref{aff72}}
\and K.~Paterson\orcid{0000-0001-8340-3486}\inst{\ref{aff84}}
\and L.~Patrizii\inst{\ref{aff34}}
\and A.~Pisani\orcid{0000-0002-6146-4437}\inst{\ref{aff73},\ref{aff178}}
\and D.~Potter\orcid{0000-0002-0757-5195}\inst{\ref{aff168}}
\and S.~Quai\orcid{0000-0002-0449-8163}\inst{\ref{aff100},\ref{aff28}}
\and M.~Radovich\orcid{0000-0002-3585-866X}\inst{\ref{aff35}}
\and P.-F.~Rocci\inst{\ref{aff25}}
\and G.~Rodighiero\orcid{0000-0002-9415-2296}\inst{\ref{aff118},\ref{aff35}}
\and S.~Sacquegna\orcid{0000-0002-8433-6630}\inst{\ref{aff152},\ref{aff153},\ref{aff154}}
\and M.~Sahl\'en\orcid{0000-0003-0973-4804}\inst{\ref{aff179}}
\and D.~B.~Sanders\orcid{0000-0002-1233-9998}\inst{\ref{aff59}}
\and E.~Sarpa\orcid{0000-0002-1256-655X}\inst{\ref{aff32},\ref{aff138},\ref{aff31}}
\and C.~Scarlata\orcid{0000-0002-9136-8876}\inst{\ref{aff16}}
\and J.~Schaye\orcid{0000-0002-0668-5560}\inst{\ref{aff51}}
\and A.~Schneider\orcid{0000-0001-7055-8104}\inst{\ref{aff168}}
\and M.~Schultheis\inst{\ref{aff102}}
\and D.~Sciotti\orcid{0009-0008-4519-2620}\inst{\ref{aff57},\ref{aff99}}
\and E.~Sellentin\inst{\ref{aff180},\ref{aff51}}
\and F.~Shankar\orcid{0000-0001-8973-5051}\inst{\ref{aff181}}
\and L.~C.~Smith\orcid{0000-0002-3259-2771}\inst{\ref{aff182}}
\and K.~Tanidis\orcid{0000-0001-9843-5130}\inst{\ref{aff13}}
\and G.~Testera\inst{\ref{aff41}}
\and R.~Teyssier\orcid{0000-0001-7689-0933}\inst{\ref{aff178}}
\and S.~Tosi\orcid{0000-0002-7275-9193}\inst{\ref{aff40},\ref{aff136}}
\and A.~Troja\orcid{0000-0003-0239-4595}\inst{\ref{aff118},\ref{aff72}}
\and M.~Tucci\inst{\ref{aff70}}
\and C.~Valieri\inst{\ref{aff34}}
\and A.~Venhola\orcid{0000-0001-6071-4564}\inst{\ref{aff183}}
\and D.~Vergani\orcid{0000-0003-0898-2216}\inst{\ref{aff28}}
\and G.~Verza\orcid{0000-0002-1886-8348}\inst{\ref{aff184}}
\and P.~Vielzeuf\orcid{0000-0003-2035-9339}\inst{\ref{aff73}}
\and N.~A.~Walton\orcid{0000-0003-3983-8778}\inst{\ref{aff182}}
\and E.~Soubrie\orcid{0000-0001-9295-1863}\inst{\ref{aff25}}
\and D.~Scott\orcid{0000-0002-6878-9840}\inst{\ref{aff185}}}
										   
\institute{David A. Dunlap Department of Astronomy \& Astrophysics, University of Toronto, 50 St George Street, Toronto, Ontario M5S 3H4, Canada\label{aff1}
\and
Jodrell Bank Centre for Astrophysics, Department of Physics and Astronomy, University of Manchester, Oxford Road, Manchester M13 9PL, UK\label{aff2}
\and
Instituto de Astrof\'{\i}sica de Canarias, V\'{\i}a L\'actea, 38205 La Laguna, Tenerife, Spain\label{aff3}
\and
Instituto de Astrof\'isica de Canarias (IAC); Departamento de Astrof\'isica, Universidad de La Laguna (ULL), 38200, La Laguna, Tenerife, Spain\label{aff4}
\and
Universit\'e PSL, Observatoire de Paris, Sorbonne Universit\'e, CNRS, LERMA, 75014, Paris, France\label{aff5}
\and
Universit\'e Paris-Cit\'e, 5 Rue Thomas Mann, 75013, Paris, France\label{aff6}
\and
Centre de Recherche Astrophysique de Lyon, UMR5574, CNRS, Universit\'e Claude Bernard Lyon 1, ENS de Lyon, 69230, Saint-Genis-Laval, France\label{aff7}
\and
Departments of Physics and Astronomy, Haverford College, 370 Lancaster Avenue, Haverford, PA 19041, USA\label{aff8}
\and
ESAC/ESA, Camino Bajo del Castillo, s/n., Urb. Villafranca del Castillo, 28692 Villanueva de la Ca\~nada, Madrid, Spain\label{aff9}
\and
School of Physical Sciences, The Open University, Milton Keynes, MK7 6AA, UK\label{aff10}
\and
INAF-Osservatorio Astronomico di Trieste, Via G. B. Tiepolo 11, 34143 Trieste, Italy\label{aff11}
\and
Centro de Astrobiolog\'ia (CAB), CSIC-INTA, ESAC Campus, Camino Bajo del Castillo s/n, 28692 Villanueva de la Ca\~nada, Madrid, Spain\label{aff12}
\and
Department of Physics, Oxford University, Keble Road, Oxford OX1 3RH, UK\label{aff13}
\and
Department of Physics, Lancaster University, Lancaster, LA1 4YB, UK\label{aff14}
\and
Masaryk University, Kotl\'{a}\v{r}sk\'{a} 2, Brno, 611 37, Czech Republic\label{aff15}
\and
Minnesota Institute for Astrophysics, University of Minnesota, 116 Church St SE, Minneapolis, MN 55455, USA\label{aff16}
\and
University of Alabama, Tuscaloosa, AL 35487, USA\label{aff17}
\and
Citizen Scientist, Zooniverse c/o University of Oxford,  Keble Road, Oxford OX1 3RH, UK\label{aff18}
\and
Institute for Theoretical Particle Physics and Cosmology (TTK), RWTH Aachen University, 52056 Aachen, Germany\label{aff19}
\and
Dipartimento di Fisica, Sapienza Universit\`a di Roma, Piazzale Aldo Moro 2, 00185 Roma, Italy\label{aff20}
\and
Instituto de F\'isica de Cantabria, Edificio Juan Jord\'a, Avenida de los Castros, 39005 Santander, Spain\label{aff21}
\and
ELTE E\"otv\"os Lor\'and University, Institute of Physics and Astronomy, P\'azm\'any P. st. 1/A, H-1171 Budapest, Hungary\label{aff22}
\and
MTA-CSFK Lend\"ulet Large-Scale Structure Research Group, Konkoly-Thege Mikl\'os \'ut 15-17, H-1121 Budapest, Hungary\label{aff23}
\and
Konkoly Observatory, HUN-REN CSFK, MTA Centre of Excellence, Budapest, Konkoly Thege Mikl\'os {\'u}t 15-17. H-1121, Hungary\label{aff24}
\and
Universit\'e Paris-Saclay, CNRS, Institut d'astrophysique spatiale, 91405, Orsay, France\label{aff25}
\and
School of Mathematics and Physics, University of Surrey, Guildford, Surrey, GU2 7XH, UK\label{aff26}
\and
INAF-Osservatorio Astronomico di Brera, Via Brera 28, 20122 Milano, Italy\label{aff27}
\and
INAF-Osservatorio di Astrofisica e Scienza dello Spazio di Bologna, Via Piero Gobetti 93/3, 40129 Bologna, Italy\label{aff28}
\and
Universit\'e Paris-Saclay, Universit\'e Paris Cit\'e, CEA, CNRS, AIM, 91191, Gif-sur-Yvette, France\label{aff29}
\and
IFPU, Institute for Fundamental Physics of the Universe, via Beirut 2, 34151 Trieste, Italy\label{aff30}
\and
INFN, Sezione di Trieste, Via Valerio 2, 34127 Trieste TS, Italy\label{aff31}
\and
SISSA, International School for Advanced Studies, Via Bonomea 265, 34136 Trieste TS, Italy\label{aff32}
\and
Dipartimento di Fisica e Astronomia, Universit\`a di Bologna, Via Gobetti 93/2, 40129 Bologna, Italy\label{aff33}
\and
INFN-Sezione di Bologna, Viale Berti Pichat 6/2, 40127 Bologna, Italy\label{aff34}
\and
INAF-Osservatorio Astronomico di Padova, Via dell'Osservatorio 5, 35122 Padova, Italy\label{aff35}
\and
Centre National d'Etudes Spatiales -- Centre spatial de Toulouse, 18 avenue Edouard Belin, 31401 Toulouse Cedex 9, France\label{aff36}
\and
Max Planck Institute for Extraterrestrial Physics, Giessenbachstr. 1, 85748 Garching, Germany\label{aff37}
\and
Universit\"ats-Sternwarte M\"unchen, Fakult\"at f\"ur Physik, Ludwig-Maximilians-Universit\"at M\"unchen, Scheinerstrasse 1, 81679 M\"unchen, Germany\label{aff38}
\and
Space Science Data Center, Italian Space Agency, via del Politecnico snc, 00133 Roma, Italy\label{aff39}
\and
Dipartimento di Fisica, Universit\`a di Genova, Via Dodecaneso 33, 16146, Genova, Italy\label{aff40}
\and
INFN-Sezione di Genova, Via Dodecaneso 33, 16146, Genova, Italy\label{aff41}
\and
Department of Physics "E. Pancini", University Federico II, Via Cinthia 6, 80126, Napoli, Italy\label{aff42}
\and
INAF-Osservatorio Astronomico di Capodimonte, Via Moiariello 16, 80131 Napoli, Italy\label{aff43}
\and
Instituto de Astrof\'isica e Ci\^encias do Espa\c{c}o, Universidade do Porto, CAUP, Rua das Estrelas, PT4150-762 Porto, Portugal\label{aff44}
\and
Faculdade de Ci\^encias da Universidade do Porto, Rua do Campo de Alegre, 4150-007 Porto, Portugal\label{aff45}
\and
Dipartimento di Fisica, Universit\`a degli Studi di Torino, Via P. Giuria 1, 10125 Torino, Italy\label{aff46}
\and
INFN-Sezione di Torino, Via P. Giuria 1, 10125 Torino, Italy\label{aff47}
\and
INAF-Osservatorio Astrofisico di Torino, Via Osservatorio 20, 10025 Pino Torinese (TO), Italy\label{aff48}
\and
European Space Agency/ESTEC, Keplerlaan 1, 2201 AZ Noordwijk, The Netherlands\label{aff49}
\and
Institute Lorentz, Leiden University, Niels Bohrweg 2, 2333 CA Leiden, The Netherlands\label{aff50}
\and
Leiden Observatory, Leiden University, Einsteinweg 55, 2333 CC Leiden, The Netherlands\label{aff51}
\and
INAF-IASF Milano, Via Alfonso Corti 12, 20133 Milano, Italy\label{aff52}
\and
Centro de Investigaciones Energ\'eticas, Medioambientales y Tecnol\'ogicas (CIEMAT), Avenida Complutense 40, 28040 Madrid, Spain\label{aff53}
\and
Port d'Informaci\'{o} Cient\'{i}fica, Campus UAB, C. Albareda s/n, 08193 Bellaterra (Barcelona), Spain\label{aff54}
\and
Institute of Space Sciences (ICE, CSIC), Campus UAB, Carrer de Can Magrans, s/n, 08193 Barcelona, Spain\label{aff55}
\and
Institut d'Estudis Espacials de Catalunya (IEEC),  Edifici RDIT, Campus UPC, 08860 Castelldefels, Barcelona, Spain\label{aff56}
\and
INAF-Osservatorio Astronomico di Roma, Via Frascati 33, 00078 Monteporzio Catone, Italy\label{aff57}
\and
INFN section of Naples, Via Cinthia 6, 80126, Napoli, Italy\label{aff58}
\and
Institute for Astronomy, University of Hawaii, 2680 Woodlawn Drive, Honolulu, HI 96822, USA\label{aff59}
\and
Dipartimento di Fisica e Astronomia "Augusto Righi" - Alma Mater Studiorum Universit\`a di Bologna, Viale Berti Pichat 6/2, 40127 Bologna, Italy\label{aff60}
\and
Institute for Astronomy, University of Edinburgh, Royal Observatory, Blackford Hill, Edinburgh EH9 3HJ, UK\label{aff61}
\and
European Space Agency/ESRIN, Largo Galileo Galilei 1, 00044 Frascati, Roma, Italy\label{aff62}
\and
Universit\'e Claude Bernard Lyon 1, CNRS/IN2P3, IP2I Lyon, UMR 5822, Villeurbanne, F-69100, France\label{aff63}
\and
Institut de Ci\`{e}ncies del Cosmos (ICCUB), Universitat de Barcelona (IEEC-UB), Mart\'{i} i Franqu\`{e}s 1, 08028 Barcelona, Spain\label{aff64}
\and
Instituci\'o Catalana de Recerca i Estudis Avan\c{c}ats (ICREA), Passeig de Llu\'{\i}s Companys 23, 08010 Barcelona, Spain\label{aff65}
\and
UCB Lyon 1, CNRS/IN2P3, IUF, IP2I Lyon, 4 rue Enrico Fermi, 69622 Villeurbanne, France\label{aff66}
\and
Mullard Space Science Laboratory, University College London, Holmbury St Mary, Dorking, Surrey RH5 6NT, UK\label{aff67}
\and
Departamento de F\'isica, Faculdade de Ci\^encias, Universidade de Lisboa, Edif\'icio C8, Campo Grande, PT1749-016 Lisboa, Portugal\label{aff68}
\and
Instituto de Astrof\'isica e Ci\^encias do Espa\c{c}o, Faculdade de Ci\^encias, Universidade de Lisboa, Campo Grande, 1749-016 Lisboa, Portugal\label{aff69}
\and
Department of Astronomy, University of Geneva, ch. d'Ecogia 16, 1290 Versoix, Switzerland\label{aff70}
\and
INAF-Istituto di Astrofisica e Planetologia Spaziali, via del Fosso del Cavaliere, 100, 00100 Roma, Italy\label{aff71}
\and
INFN-Padova, Via Marzolo 8, 35131 Padova, Italy\label{aff72}
\and
Aix-Marseille Universit\'e, CNRS/IN2P3, CPPM, Marseille, France\label{aff73}
\and
INFN-Bologna, Via Irnerio 46, 40126 Bologna, Italy\label{aff74}
\and
School of Physics, HH Wills Physics Laboratory, University of Bristol, Tyndall Avenue, Bristol, BS8 1TL, UK\label{aff75}
\and
FRACTAL S.L.N.E., calle Tulip\'an 2, Portal 13 1A, 28231, Las Rozas de Madrid, Spain\label{aff76}
\and
NRC Herzberg, 5071 West Saanich Rd, Victoria, BC V9E 2E7, Canada\label{aff77}
\and
Institute of Theoretical Astrophysics, University of Oslo, P.O. Box 1029 Blindern, 0315 Oslo, Norway\label{aff78}
\and
Jet Propulsion Laboratory, California Institute of Technology, 4800 Oak Grove Drive, Pasadena, CA, 91109, USA\label{aff79}
\and
Felix Hormuth Engineering, Goethestr. 17, 69181 Leimen, Germany\label{aff80}
\and
Technical University of Denmark, Elektrovej 327, 2800 Kgs. Lyngby, Denmark\label{aff81}
\and
Cosmic Dawn Center (DAWN), Denmark\label{aff82}
\and
Institut d'Astrophysique de Paris, UMR 7095, CNRS, and Sorbonne Universit\'e, 98 bis boulevard Arago, 75014 Paris, France\label{aff83}
\and
Max-Planck-Institut f\"ur Astronomie, K\"onigstuhl 17, 69117 Heidelberg, Germany\label{aff84}
\and
NASA Goddard Space Flight Center, Greenbelt, MD 20771, USA\label{aff85}
\and
Department of Physics and Astronomy, University College London, Gower Street, London WC1E 6BT, UK\label{aff86}
\and
Department of Physics and Helsinki Institute of Physics, Gustaf H\"allstr\"omin katu 2, 00014 University of Helsinki, Finland\label{aff87}
\and
Universit\'e de Gen\`eve, D\'epartement de Physique Th\'eorique and Centre for Astroparticle Physics, 24 quai Ernest-Ansermet, CH-1211 Gen\`eve 4, Switzerland\label{aff88}
\and
Department of Physics, P.O. Box 64, 00014 University of Helsinki, Finland\label{aff89}
\and
Helsinki Institute of Physics, Gustaf H{\"a}llstr{\"o}min katu 2, University of Helsinki, Helsinki, Finland\label{aff90}
\and
Centre de Calcul de l'IN2P3/CNRS, 21 avenue Pierre de Coubertin 69627 Villeurbanne Cedex, France\label{aff91}
\and
Laboratoire d'etude de l'Univers et des phenomenes eXtremes, Observatoire de Paris, Universit\'e PSL, Sorbonne Universit\'e, CNRS, 92190 Meudon, France\label{aff92}
\and
Aix-Marseille Universit\'e, CNRS, CNES, LAM, Marseille, France\label{aff93}
\and
SKA Observatory, Jodrell Bank, Lower Withington, Macclesfield, Cheshire SK11 9FT, UK\label{aff94}
\and
Dipartimento di Fisica "Aldo Pontremoli", Universit\`a degli Studi di Milano, Via Celoria 16, 20133 Milano, Italy\label{aff95}
\and
INFN-Sezione di Milano, Via Celoria 16, 20133 Milano, Italy\label{aff96}
\and
University of Applied Sciences and Arts of Northwestern Switzerland, School of Computer Science, 5210 Windisch, Switzerland\label{aff97}
\and
Universit\"at Bonn, Argelander-Institut f\"ur Astronomie, Auf dem H\"ugel 71, 53121 Bonn, Germany\label{aff98}
\and
INFN-Sezione di Roma, Piazzale Aldo Moro, 2 - c/o Dipartimento di Fisica, Edificio G. Marconi, 00185 Roma, Italy\label{aff99}
\and
Dipartimento di Fisica e Astronomia "Augusto Righi" - Alma Mater Studiorum Universit\`a di Bologna, via Piero Gobetti 93/2, 40129 Bologna, Italy\label{aff100}
\and
Department of Physics, Institute for Computational Cosmology, Durham University, South Road, Durham, DH1 3LE, UK\label{aff101}
\and
Universit\'e C\^{o}te d'Azur, Observatoire de la C\^{o}te d'Azur, CNRS, Laboratoire Lagrange, Bd de l'Observatoire, CS 34229, 06304 Nice cedex 4, France\label{aff102}
\and
Universit\'e Paris Cit\'e, CNRS, Astroparticule et Cosmologie, 75013 Paris, France\label{aff103}
\and
CNRS-UCB International Research Laboratory, Centre Pierre Binetruy, IRL2007, CPB-IN2P3, Berkeley, USA\label{aff104}
\and
University of Applied Sciences and Arts of Northwestern Switzerland, School of Engineering, 5210 Windisch, Switzerland\label{aff105}
\and
Institut d'Astrophysique de Paris, 98bis Boulevard Arago, 75014, Paris, France\label{aff106}
\and
Institute of Physics, Laboratory of Astrophysics, Ecole Polytechnique F\'ed\'erale de Lausanne (EPFL), Observatoire de Sauverny, 1290 Versoix, Switzerland\label{aff107}
\and
Aurora Technology for European Space Agency (ESA), Camino bajo del Castillo, s/n, Urbanizacion Villafranca del Castillo, Villanueva de la Ca\~nada, 28692 Madrid, Spain\label{aff108}
\and
Institut de F\'{i}sica d'Altes Energies (IFAE), The Barcelona Institute of Science and Technology, Campus UAB, 08193 Bellaterra (Barcelona), Spain\label{aff109}
\and
School of Mathematics, Statistics and Physics, Newcastle University, Herschel Building, Newcastle-upon-Tyne, NE1 7RU, UK\label{aff110}
\and
DARK, Niels Bohr Institute, University of Copenhagen, Jagtvej 155, 2200 Copenhagen, Denmark\label{aff111}
\and
Waterloo Centre for Astrophysics, University of Waterloo, Waterloo, Ontario N2L 3G1, Canada\label{aff112}
\and
Department of Physics and Astronomy, University of Waterloo, Waterloo, Ontario N2L 3G1, Canada\label{aff113}
\and
Perimeter Institute for Theoretical Physics, Waterloo, Ontario N2L 2Y5, Canada\label{aff114}
\and
Institute of Space Science, Str. Atomistilor, nr. 409 M\u{a}gurele, Ilfov, 077125, Romania\label{aff115}
\and
Consejo Superior de Investigaciones Cientificas, Calle Serrano 117, 28006 Madrid, Spain\label{aff116}
\and
Universidad de La Laguna, Departamento de Astrof\'{\i}sica, 38206 La Laguna, Tenerife, Spain\label{aff117}
\and
Dipartimento di Fisica e Astronomia "G. Galilei", Universit\`a di Padova, Via Marzolo 8, 35131 Padova, Italy\label{aff118}
\and
Caltech/IPAC, 1200 E. California Blvd., Pasadena, CA 91125, USA\label{aff119}
\and
Institut f\"ur Theoretische Physik, University of Heidelberg, Philosophenweg 16, 69120 Heidelberg, Germany\label{aff120}
\and
Institut de Recherche en Astrophysique et Plan\'etologie (IRAP), Universit\'e de Toulouse, CNRS, UPS, CNES, 14 Av. Edouard Belin, 31400 Toulouse, France\label{aff121}
\and
Universit\'e St Joseph; Faculty of Sciences, Beirut, Lebanon\label{aff122}
\and
Departamento de F\'isica, FCFM, Universidad de Chile, Blanco Encalada 2008, Santiago, Chile\label{aff123}
\and
Universit\"at Innsbruck, Institut f\"ur Astro- und Teilchenphysik, Technikerstr. 25/8, 6020 Innsbruck, Austria\label{aff124}
\and
Satlantis, University Science Park, Sede Bld 48940, Leioa-Bilbao, Spain\label{aff125}
\and
Infrared Processing and Analysis Center, California Institute of Technology, Pasadena, CA 91125, USA\label{aff126}
\and
Instituto de Astrof\'isica e Ci\^encias do Espa\c{c}o, Faculdade de Ci\^encias, Universidade de Lisboa, Tapada da Ajuda, 1349-018 Lisboa, Portugal\label{aff127}
\and
Cosmic Dawn Center (DAWN)\label{aff128}
\and
Niels Bohr Institute, University of Copenhagen, Jagtvej 128, 2200 Copenhagen, Denmark\label{aff129}
\and
Universidad Polit\'ecnica de Cartagena, Departamento de Electr\'onica y Tecnolog\'ia de Computadoras,  Plaza del Hospital 1, 30202 Cartagena, Spain\label{aff130}
\and
Kapteyn Astronomical Institute, University of Groningen, PO Box 800, 9700 AV Groningen, The Netherlands\label{aff131}
\and
Dipartimento di Fisica e Scienze della Terra, Universit\`a degli Studi di Ferrara, Via Giuseppe Saragat 1, 44122 Ferrara, Italy\label{aff132}
\and
Istituto Nazionale di Fisica Nucleare, Sezione di Ferrara, Via Giuseppe Saragat 1, 44122 Ferrara, Italy\label{aff133}
\and
INAF, Istituto di Radioastronomia, Via Piero Gobetti 101, 40129 Bologna, Italy\label{aff134}
\and
INAF - Osservatorio Astronomico di Brera, via Emilio Bianchi 46, 23807 Merate, Italy\label{aff135}
\and
INAF-Osservatorio Astronomico di Brera, Via Brera 28, 20122 Milano, Italy, and INFN-Sezione di Genova, Via Dodecaneso 33, 16146, Genova, Italy\label{aff136}
\and
ICL, Junia, Universit\'e Catholique de Lille, LITL, 59000 Lille, France\label{aff137}
\and
ICSC - Centro Nazionale di Ricerca in High Performance Computing, Big Data e Quantum Computing, Via Magnanelli 2, Bologna, Italy\label{aff138}
\and
Instituto de F\'isica Te\'orica UAM-CSIC, Campus de Cantoblanco, 28049 Madrid, Spain\label{aff139}
\and
CERCA/ISO, Department of Physics, Case Western Reserve University, 10900 Euclid Avenue, Cleveland, OH 44106, USA\label{aff140}
\and
Technical University of Munich, TUM School of Natural Sciences, Physics Department, James-Franck-Str.~1, 85748 Garching, Germany\label{aff141}
\and
Max-Planck-Institut f\"ur Astrophysik, Karl-Schwarzschild-Str.~1, 85748 Garching, Germany\label{aff142}
\and
Laboratoire Univers et Th\'eorie, Observatoire de Paris, Universit\'e PSL, Universit\'e Paris Cit\'e, CNRS, 92190 Meudon, France\label{aff143}
\and
Departamento de F{\'\i}sica Fundamental. Universidad de Salamanca. Plaza de la Merced s/n. 37008 Salamanca, Spain\label{aff144}
\and
Universit\'e de Strasbourg, CNRS, Observatoire astronomique de Strasbourg, UMR 7550, 67000 Strasbourg, France\label{aff145}
\and
Center for Data-Driven Discovery, Kavli IPMU (WPI), UTIAS, The University of Tokyo, Kashiwa, Chiba 277-8583, Japan\label{aff146}
\and
Ludwig-Maximilians-University, Schellingstrasse 4, 80799 Munich, Germany\label{aff147}
\and
Max-Planck-Institut f\"ur Physik, Boltzmannstr. 8, 85748 Garching, Germany\label{aff148}
\and
California Institute of Technology, 1200 E California Blvd, Pasadena, CA 91125, USA\label{aff149}
\and
Departamento de F\'{i}sica Te\'{o}rica, At\'{o}mica y \'{O}ptica, Universidad de Valladolid, 47011 Valladolid, Spain\label{aff150}
\and
Department of Physics \& Astronomy, University of California Irvine, Irvine CA 92697, USA\label{aff151}
\and
Department of Mathematics and Physics E. De Giorgi, University of Salento, Via per Arnesano, CP-I93, 73100, Lecce, Italy\label{aff152}
\and
INFN, Sezione di Lecce, Via per Arnesano, CP-193, 73100, Lecce, Italy\label{aff153}
\and
INAF-Sezione di Lecce, c/o Dipartimento Matematica e Fisica, Via per Arnesano, 73100, Lecce, Italy\label{aff154}
\and
Departamento F\'isica Aplicada, Universidad Polit\'ecnica de Cartagena, Campus Muralla del Mar, 30202 Cartagena, Murcia, Spain\label{aff155}
\and
CEA Saclay, DFR/IRFU, Service d'Astrophysique, Bat. 709, 91191 Gif-sur-Yvette, France\label{aff156}
\and
Institute of Cosmology and Gravitation, University of Portsmouth, Portsmouth PO1 3FX, UK\label{aff157}
\and
Department of Computer Science, Aalto University, PO Box 15400, Espoo, FI-00 076, Finland\label{aff158}
\and
Instituto de Astrof\'\i sica de Canarias, c/ Via Lactea s/n, La Laguna 38200, Spain. Departamento de Astrof\'\i sica de la Universidad de La Laguna, Avda. Francisco Sanchez, La Laguna, 38200, Spain\label{aff159}
\and
Ruhr University Bochum, Faculty of Physics and Astronomy, Astronomical Institute (AIRUB), German Centre for Cosmological Lensing (GCCL), 44780 Bochum, Germany\label{aff160}
\and
Department of Physics and Astronomy, Vesilinnantie 5, 20014 University of Turku, Finland\label{aff161}
\and
Serco for European Space Agency (ESA), Camino bajo del Castillo, s/n, Urbanizacion Villafranca del Castillo, Villanueva de la Ca\~nada, 28692 Madrid, Spain\label{aff162}
\and
ARC Centre of Excellence for Dark Matter Particle Physics, Melbourne, Australia\label{aff163}
\and
Centre for Astrophysics \& Supercomputing, Swinburne University of Technology,  Hawthorn, Victoria 3122, Australia\label{aff164}
\and
Department of Physics and Astronomy, University of the Western Cape, Bellville, Cape Town, 7535, South Africa\label{aff165}
\and
DAMTP, Centre for Mathematical Sciences, Wilberforce Road, Cambridge CB3 0WA, UK\label{aff166}
\and
Kavli Institute for Cosmology Cambridge, Madingley Road, Cambridge, CB3 0HA, UK\label{aff167}
\and
Department of Astrophysics, University of Zurich, Winterthurerstrasse 190, 8057 Zurich, Switzerland\label{aff168}
\and
Department of Physics, Centre for Extragalactic Astronomy, Durham University, South Road, Durham, DH1 3LE, UK\label{aff169}
\and
IRFU, CEA, Universit\'e Paris-Saclay 91191 Gif-sur-Yvette Cedex, France\label{aff170}
\and
Oskar Klein Centre for Cosmoparticle Physics, Department of Physics, Stockholm University, Stockholm, SE-106 91, Sweden\label{aff171}
\and
Astrophysics Group, Blackett Laboratory, Imperial College London, London SW7 2AZ, UK\label{aff172}
\and
Univ. Grenoble Alpes, CNRS, Grenoble INP, LPSC-IN2P3, 53, Avenue des Martyrs, 38000, Grenoble, France\label{aff173}
\and
INAF-Osservatorio Astrofisico di Arcetri, Largo E. Fermi 5, 50125, Firenze, Italy\label{aff174}
\and
Centro de Astrof\'{\i}sica da Universidade do Porto, Rua das Estrelas, 4150-762 Porto, Portugal\label{aff175}
\and
HE Space for European Space Agency (ESA), Camino bajo del Castillo, s/n, Urbanizacion Villafranca del Castillo, Villanueva de la Ca\~nada, 28692 Madrid, Spain\label{aff176}
\and
Dipartimento di Fisica - Sezione di Astronomia, Universit\`a di Trieste, Via Tiepolo 11, 34131 Trieste, Italy\label{aff177}
\and
Department of Astrophysical Sciences, Peyton Hall, Princeton University, Princeton, NJ 08544, USA\label{aff178}
\and
Theoretical astrophysics, Department of Physics and Astronomy, Uppsala University, Box 515, 751 20 Uppsala, Sweden\label{aff179}
\and
Mathematical Institute, University of Leiden, Einsteinweg 55, 2333 CA Leiden, The Netherlands\label{aff180}
\and
School of Physics \& Astronomy, University of Southampton, Highfield Campus, Southampton SO17 1BJ, UK\label{aff181}
\and
Institute of Astronomy, University of Cambridge, Madingley Road, Cambridge CB3 0HA, UK\label{aff182}
\and
Space physics and astronomy research unit, University of Oulu, Pentti Kaiteran katu 1, FI-90014 Oulu, Finland\label{aff183}
\and
Center for Computational Astrophysics, Flatiron Institute, 162 5th Avenue, 10010, New York, NY, USA\label{aff184}
\and
Department of Physics and Astronomy, University of British Columbia, Vancouver, BC V6T 1Z1, Canada\label{aff185}}    

\abstract{
We present a detailed visual morphology catalogue for \Euclid's Quick Release 1 (Q1). Our catalogue includes galaxy features such as bars, spiral arms, and ongoing mergers, for the 378\,000 bright ($\IE < 20.5$) or extended (area $\geq 700\,$pixels) galaxies in Q1. The catalogue was created by finetuning the \texttt{Zoobot} galaxy foundation models on annotations from an intensive one month campaign by Galaxy Zoo volunteers. Our measurements are fully automated and hence fully scaleable. This catalogue is the first 0.4\% of the approximately 100 million galaxies where \Euclid will ultimately resolve detailed morphology.
}
%
%
%

   \titlerunning{\Euclid Q1: First visual morphology catalogue}
   \authorrunning{Euclid Collaboration: M.~Walmsley et. al.}
   
   \maketitle
%
%
%

\section{\label{sc:Intro}Introduction}
 
Detailed visual morphology refers to the recognizable features which comprise a galaxy, such as bars, spiral arms, and tidal tails \citep{hubble_extragalactic_1926,de_vaucouleurs_classification_1959,Toomre1972,sellwoodSpiralsGalaxies2022}. Understanding how galaxies acquire their stellar structure provides key insights into the processes driving mass assembly in the Universe (e.g.~\citealp{2011ApJ...742...96W,2015Sci...348..314T,2016MNRAS.462.4495H})  Visual morphology has historically also described the method of detection; we measure these features visually, by eye. Those eyes may either belong to professional astronomers \citep{Nair2010,2011A&A...532A..74B,Buta2015} or to members of the public taking part in citizen science projects such as Galaxy Zoo \citep{Lintott2008, Masters2019a} and Galaxy Cruise \citep{tanaka_galaxy_2023}. Visual morphology complements parametric morphology, such as S\'ersic fitting \citep{sersic_influence_1963}, and non-parametric morphology, such as concentration and asymmetry \citep{morgan_preliminary_1958,Conselice2000,shimasaku_statistical_2001,Abraham2003a}, which both use rule-based automated methods to interpret galaxy images. Parametric and non-parametric morphology have historically been together known as `quantitative' morphology, contrasting with `qualitative' visual morphology.

The complexity of galaxies is greater than the complexity we are able to express in code.
Galaxies have features which are too complex for our rule-based methods, but are real nonetheless (see e.g., \citealt{lintott_galaxy_2009,Rudnick2021,bowles_radio_2023,gordonUncoveringTidalTreasures2024}). Astronomers have therefore faced a trade-off. One can use visual morphology to capture detailed features, or quantitive morphology to make measurements which are scaleable and reproducible \citep{conseliceEvolutionGalaxyStructure2014}. There is also a spectrum of work between these two extremes that makes detailed automated measurements under a degree of manual supervision and tuning, e.g., \texttt{galfit} \citep{peng_detailed_2002} and Galaxy Zoo Builder \citep{Lingard2020}.

Recent advances in computer vision make it possible, even straightforward, to automate some visual judgements. 
Seminal work by \cite{Dieleman2015} won the Galaxy Challenge, a Kaggle competition to predict the visual judgements of Galaxy Zoo volunteers, and in doing so introduced deep learning to astronomy.
A decade later, deep learning is a ubiquitous tool for measuring visual morphology (e.g., \citealt{Khan2018,Abraham2018,Pearson2019Mergers,Ghosh2020,Bom2021a,ciprijanovic_semi-supervised_2022} and review by~\citealt{HuertasCompany2023Dawes}).
Citizen science and deep learning have together underpinned detailed visual morphology catalogues for the \textit{Hubble} Space Telescope \citep{Huertas-Company2015a}, the Sloan Digital Sky Survey \citep{Sanchez2018}, the Dark Energy Camera Legacy Survey \citep{Walmsley2022decals}, and the companion Legacy Surveys \citep{walmsleyGalaxyZooDESI2023,yeGalaxyZooDECaLS2025}.

Our core advance here is timing. 
Morphology catalogues typically follow years after a telescope data release -- 3.5 to 5 years for each of Galaxy Zoo's morphology catalogues, for example.
Much of this time is needed for volunteers to annotate galaxies and, more recently, to train models.

What if we made morphology measurements at the same time as the survey takes images, just as we already do for other automated measurements?
Placing a trained deep learning model within the survey image processing pipeline allows for immediate morphology measurements and immediate use by scientists. 
Our models can be trained quickly because we use new `foundation' models (described in Sect. \ref{sec:finetuning}) that need fewer examples to learn to classify new surveys. In this work, we deliver a detailed visual morphology catalogue for \Euclid in weeks instead of years.

\Euclid will resolve the detailed visual morphology of at least an order of magnitude more galaxies than have ever been measured. The largest current detailed morphology catalogues use images from the DESI Legacy Surveys \citep{walmsleyGalaxyZooDESI2023}, with 19\,000 deg$^2$ of imaging at \ang{;;1.1} seeing; and the Sloan Digital Sky Survey \citep{Willett2013,Sanchez2018}, with 9000 deg$^2$ of imaging at \ang{;;1.3} seeing (DR7, \citealt{Abazajian2009}). The EWS will cover approximately 14\,000 deg$^2$ with a spatial resolution of $\ang{;;0.16}$ \citep{EuclidSkyVIS} -- a comparable area at ten times higher resolution. The final \Euclid morphology catalogues will include approximately $10^8$ galaxies. Here, we measure detailed morphology in the first 0.4\% -- Euclid Quick Release 1 (Q1, \citealt{Q1cite}). 

\begin{figure}
    \centering
    \includegraphics[width=\linewidth]{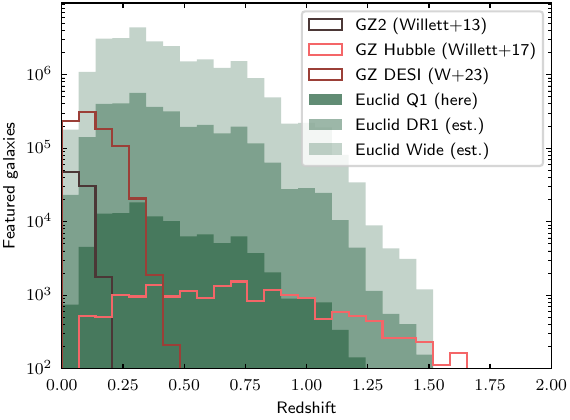}
    \caption{Galaxies with well-resolved features as a function of redshift, for morphology catalogues from \Euclid vs. Galaxy Zoo 2 (main sample, \citealt{Willett2013}), Galaxy Zoo Hubble \citep{Willett2017a}, and Galaxy Zoo DESI \citep{walmsleyGalaxyZooDESI2023}. \Euclid will outscale all current visual morphology catalogues for all redshifts below $z < 1.5$, typically by several orders of magnitude. Q1 morphologies may be especially valuable between $0.3 < z < 0.7$. We define `featured' as a galaxy with an expected volunteer vote fraction for `Featured' above 0.5 (see Sect. \ref{sec:data_access}). `Euclid Q1' is this catalogue; estimates for `Euclid DR1' and `Euclid Wide' (EWS) are made by trivially multiplying our catalogue counts by area. \Euclid redshifts are photometric \citep{Q1-TP005}.}
    \label{fig:survey_comparison}
\end{figure}

\Euclid connects low-redshift ground-based morphology measurements with high-redshift space-based measurements, enabling a continuous view of galaxy morphology through time. Figure \ref{fig:survey_comparison} compares, as a function of redshift, the number of galaxies with visual features in our Q1 catalogue vs. previous catalogues made with the Sloan Digital Sky Survey \citep{Willett2013}, the \textit{Hubble} Space Telescope (HST, \citealt{Willett2017a}), and the Legacy Surveys \citep{walmsleyGalaxyZooDESI2023}. Q1 adds an order-of-magnitude more galaxies between $0.3 < z < 0.7$. 
Straightforwardly multiplying our results by area, the EWS will ultimately increase the number of galaxies with measured morphology features between $0.3 < z < 0.7$ by around three orders of magnitude.

Our Q1 catalogue is available in two forms. First, our initial trained model is part of the \Euclid pipeline, and so the morphology measurements from that model are reported as part of the Q1 data release~\citep{Q1-TP004}. Those measurements are accessible through the ESA Science Archive Service as with other core measurements such as photometry, redshifts, and so forth. We refer to this as the \emph{pipeline catalogue}. Second, we created a separate catalogue by applying our next generation of models directly to the \Euclid images, outside of the \Euclid pipeline.
We did this to use the best possible models (which are updated more frequently than is practical within the \Euclid pipeline) and to create and share our embeddings (vectors which mathematically summarise the visual features of each galaxy). We refer to this as the \emph{dynamic catalogue}.

Our catalogue complements parallel work by \citet{Q1-TP004} and \citet{Q1-SP040} to create a morphology catalogue for Q1 with parametric and non-parametric measurements. We recommend using these traditional measurements for galaxies less extended than around 700 pixels in segmentation area\footnote{As measured by SourceExtractor++ within the MERge pipeline and reported as `SEGMENTATION\_AREA’, \citep{Q1-TP004, Bertin1996}. This roughly corresponds to \ang{;;1.5} in radius. MERge mosaic images have a pixel scale of \ang{;;0.1} per pixel.}, below which \Euclid cannot reliably resolve detailed features. \citet{Q1-SP040} includes a comparison of disk and bulge measurements using this detailed morphology catalogue and using S\'ersic fits and finds consistent results. 

Our catalogue was made possible by the efforts of 9976 Galaxy Zoo volunteers who together contributed 2.9M annotations to adapt the \texttt{Zoobot} foundation deep learning models for \Euclid images. 
These measurements, combined with parallel work using the \texttt{Zoobot} models to find strong lenses \citep{Q1-SP048,Q1-SP052, Q1-SP053, Q1-SP054, Q1-SP059}, stellar bars~\citep{Q1-SP043}, mergers \citep{Q1-SP013} and AGN \citep{Q1-SP015} demonstrate the practical value of foundation models in astronomy.

In Sect. \ref{sec:data}, we describe our selection function and image processing choices. 
In Sect. \ref{sec:citizen_science}, we describe how Galaxy Zoo volunteers contributed annotations.
In Sect. \ref{sec:finetuning}, we motivate our use of foundation models and detail the finetuning process. 
In Sect. \ref{sec:results}, we validate the performance of our finetuned models.
In Sect. \ref{sec:data_access}, we share our dynamic catalogue, embeddings, and images, and provide practical guidance on how these might be used. They can be downloaded from Zenodo\footnote{\url{https://doi.org/10.5281/zenodo.15002907}} and HuggingFace\footnote{\url{https://huggingface.co/collections/mwalmsley/euclid-67cf5a80e2a93f09e6e4df2c}}.

\section{\label{sec:data}  Data}

\subsection{\label{coverage} Coverage}

\Euclid will detect approximately 1.5 billion sources \citep{Bretonniere-EP13,EuclidSkyOverview}. The largest sources will be revealed in exquisite detail \citep{ERONearbyGals}. Most will be barely resolved. In between will be a middle ground of sources which show some suggestion of detailed morphology (the trace of a disc, an arm, a bar, etc.). When choosing which galaxies to measure for detailed morphology, where should we draw the line?

The human annotations guiding the models that, in turn, create our catalogue, come from Galaxy Zoo volunteers -- members of the public contributing their time to click through galaxy images \citep{Masters2019a}. We need to make the best possible use of Galaxy Zoo volunteers' time, particularly during the one month labelling campaign to produce the pipeline models (Sect. \ref{sec:citizen_science}). We should especially avoid showing a high ratio of featureless galaxies (`blobs') as these are relatively straightforward to classify automatically and may dissuade volunteers. Therefore, we chose the following conservative cut to select galaxies with a moderate chance of showing detailed features.

\begin{flushleft}
    \texttt{segmentation\_area} $> 1200$ pixels\\
    \texttt{OR} \hfill \break
    $\IE < 20.5 $\texttt{ AND segmentation\_area}$ > 200$ pixels \hfill \break
\end{flushleft}

We found segmentation area (the total number of pixels within the segmentation source mask from \texttt{SourceExtractor++}, as calculated by \citealt{Q1-TP004}) to be the critical factor in determining if a galaxy was well-resolved. Segmentation area is a natural proxy for assessing if a galaxy is well-resolved because each morphological feature requires sampling by some number of point spread function full-width-half-maximum (FWHM) to be resolved, and this sampling happens in two dimensions. Results using radii were broadly similar but suffered from orientation effects or asymmetric sources.
Our choice of 1200 pixels was a subjective choice with the aim of creating an engaging sample for Galaxy Zoo volunteers (see above), and was ultimately later revised for the dynamic catalogue (below).
The magnitude cut follows from the common science requirement for completeness, and is complemented by an alternative (far more generous) segmentation cut to remove galaxies where detailed features are plainly unmeasurable. Overall, this selection cut includes the brightest and most extended 0.8\% of galaxies in Q1 (\num{195716} galaxies). These form the selection shown to volunteers and measured by the pipeline models. 

For the dynamic catalogue, we reduce the \texttt{segmentation\_area} cut from $1200$ pixels to $700$ pixels (for a total of \num{380\,111} galaxies, 1.5\% of Q1). This adds \num{184\,395} galaxies which are fainter and less extended but may still have resolvable features. We do not (currently) show these less extended galaxies to Galaxy Zoo volunteers, and instead rely on our trained models to extrapolate to this regime. The dynamic catalogue includes the column `in\_extrapolated\_selection' for users to include or exclude these additional galaxies as desired. Lacking ground truth labels, we cannot make any performance claim for these galaxies, but our expert visual inspection qualitatively suggests the models continue to work similarly well -- perhaps because the images are less detailed and therefore present a less challenging computer vision task, because the segmentation area is imprecisely measured, or because the models were pretrained on similar images from other surveys \citep{walmsleyScalingLawsGalaxy2024}.  

\begin{figure}
    \centering
    \includegraphics[width=\linewidth]{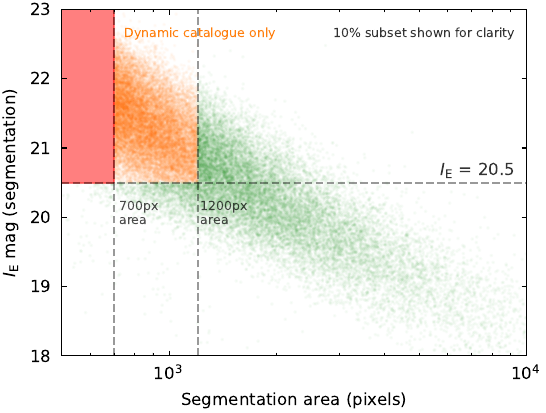}
    \caption{Cuts applied to select galaxies with resolvable visual morphology. The pipeline catalogue includes all galaxies with area > 1200 pixels or \IE < 20.5 (and area > 200 pixels). The dynamic catalogue reduces the area limit to 700 pixels. The total number of galaxies in the dynamic catalogue only (orange) vs. in both catalogues (green) are roughly equal (\num{195716} vs. \num{184395}, respectively)}
    \label{fig:selection-cuts}
\end{figure}

While we could make automated measurements of every source in Q1, visually inspecting example images suggested that galaxies with a segmentation area below around 700 pixels are insufficiently resolved to clearly show detailed features, and so we select our lowest area cut as 700 pixels and defer deep learning morphology measurements of smaller galaxies to future work.
Figure \ref{fig:selection-cuts} illustrates our choice of selection cuts.

For both selections, we additionally require \texttt{vis\_det}$=1$ and \texttt{spurious\_prob}$ < 0.2$, to remove artifacts, and require no \textit{Gaia} cross-match to remove stars. 

\subsection{\label{sec:image_processing} Image processing}

\begin{figure}
    \centering
    \includegraphics[width=\linewidth]{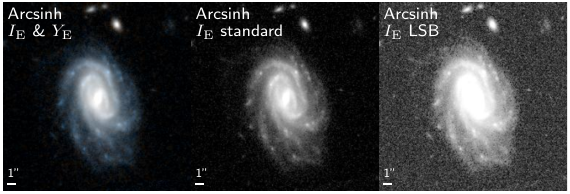}
    \caption{Example cutouts shown to volunteers. Left to right: \IE/\YE composite, \IE-only greyscale with standard processing, \IE-only greyscale with enhanced low-surface-brightness processing.}
    \label{fig:cutout_examples}
\end{figure}

We create three jpg cutouts from each source. Figure \ref{fig:cutout_examples} shows examples. The three cutouts are:
\begin{enumerate}
    \item A composite RGB image where the R channel is \YE, the B channel is \IE, and the G channel is the mean of the pixelwise flux in the other two channels, following a 99.85th percentile clip and an arcsinh stretch, i.e., $x' = $~arcsinh$(Qx)$ with $Q=100$ where $x$ is the flux in each pixel. 
    \item A greyscale image where the single channel is identical to the \IE/B channel above, maximising resolution
    \item A greyscale image where the single channel is again from \IE, but adjusted to highlight low-surface brightness features. We use the recipe from \citet{gordonUncoveringTidalTreasures2024} with a stretch of 20 and a power of 0.5, and add a 98th percentile clip.
\end{enumerate}

We designed these processing options to create a complementary set of images for volunteers; a colour image showing the general galaxy features, a maximum-sharpness (but greyscale) image, and an image aimed at highlighting low-surface brightness features which are better revealed when shown on a separate scale to the bright galaxy core. We used \YE for the colour image as \YE is the sharpest (lowest PSF FWHM) NISP band. We combined data from different instruments using the aligned and resampled mosaics provided by the MERge pipeline \citep{Q1-TP004}.

Volunteers were shown all three images in a flipbook format with the order above. We then use their responses to adapt our models. The pipeline catalogue is made by a model shown the standard \IE image (only \IE is available). The dynamic catalogue is made by a model shown the composite \IE/\YE image. 

\section{Methods\label{sec:methods}}

\subsection{\label{sec:citizen_science}Citizen science}

We presented \Euclid images to Galaxy Zoo volunteers. 
Volunteers annotated images from the Euclid Wide Survey (EWS), and \textit{not} from Q1. Our pipeline models run within the 
\Euclid pipeline that produced the Q1 data release, and so the pipeline models needed to be ready before Q1 was available. 
We showed these EWS images with permission from ESA and via a Memorandum of Understanding between the Euclid Consortium and the Zooniverse. This MoU created a framework for the Galaxy Zoo team to work with \Euclid scientists to share a small set of EWS images with the public. These images are ideal for training models that work well on the EWS, and therefore on the vast majority of galaxies \Euclid will image. We plan on returning to specifically annotate the Euclid Deep fields (including the Q1 area) once full-depth data is available.

The \Euclid survey images are available as mosaic tiles of $\ang{;32;}\times\ang{;32;}$ \citep{Q1-TP004}.
Galaxies were selected from a set of tiles spread uniformly across the EWS area\footnote{We selected tiles by picking a random tile, then picking the most distant tile to all previous tiles, repeatedly.}. All tiles were drawn from the southern half of the EWS (declination $< 0$) as source catalogue data were not yet available for the northern half.

9976 volunteers contributed 2.9 million annotations of \num{114000} galaxies. Of those, 1.56 million annotations were made in the initial one month labelling campaign and used to train the pipeline model. All 2.9 million annotations were used to train the dynamic catalogue model.

\begin{figure}
    \centering
    \includegraphics[width=\linewidth]{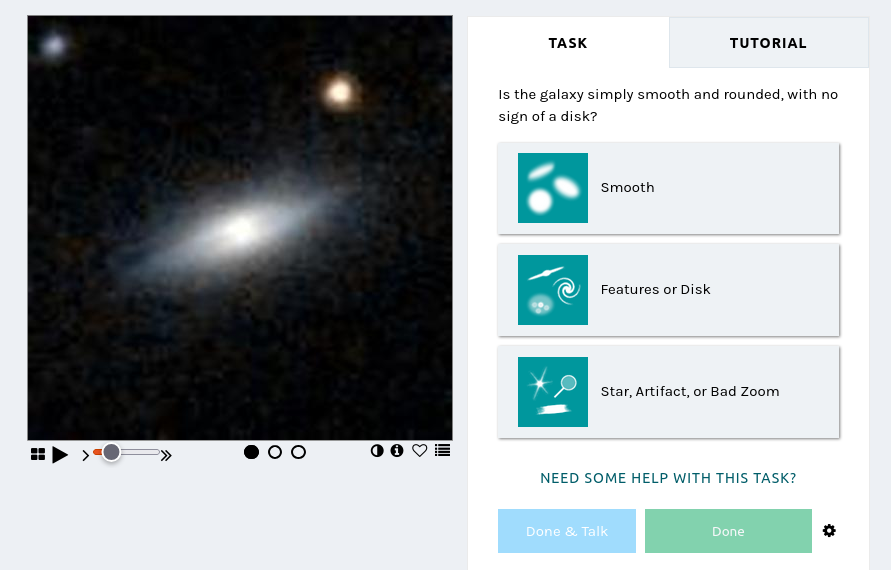}
    \caption{Galaxy Zoo annotation interface, as shown to volunteers. Volunteers answer a tree of questions, where the next question depends on the previous answer; only the first question is shown here.}
    \label{fig:galaxy_zoo}
\end{figure}

\begin{figure}
    \centering
    \includegraphics[width=\linewidth]{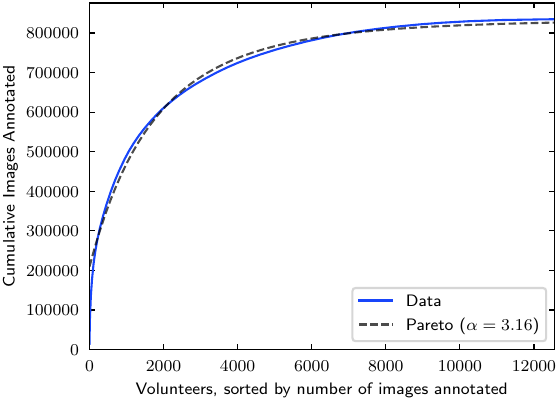}
    \caption{Cumulative images annotated when selecting the $N$ volunteers who annotated the most images. The bulk of annotations come from highly-engaged volunteers.}
    \label{fig:pareto}
\end{figure}

Volunteers were presented with a pop-up tutorial, and shown examples in `help' instructions for each question alongside a site-wide `field guide'. The annotation interface is shown in Fig. \ref{fig:galaxy_zoo}.
In line with previous Galaxy Zoo projects, a small portion of highly-engaged volunteers contribute the bulk of the annotations. Volunteer contributions are well-modelled by a Pareto distribution (Fig. \ref{fig:pareto}).

We decided to request five volunteer annotations per galaxy for the most galaxies (\num{110000} of \num{114000}). Five volunteers is far fewer than typical; Galaxy Zoo has historically collected classifications from 40 volunteers per galaxy (for example \citealt{Willett2013}). Asking fewer volunteers per galaxy increases the noise in our labels but also increases the diversity of galaxies labelled. We hypothesize that this is a useful trade-off for maximising model performance provided the model loss function can handle uncertain labels (see Sect. \ref{sec:finetuning}). To accurately measure model performance, we also chose a small random subset (\num{3500}) to be annotated by 25 volunteers.

Maximising model performance is our key goal because only a tiny fraction of galaxies imaged by \Euclid will ever be seen by humans. The complete EWS will include approximately $10^8$ galaxies passing our selection cuts above, compared to $\lesssim 2$M galaxies in all Galaxy Zoo projects over the last 15 years.
It is impossible for volunteers to annotate more than a few percent of \Euclid galaxies.
We therefore do not expect scientists to use the volunteer annotations directly, as in all Galaxy Zoo projects prior to \citet{Walmsley2022decals}, but instead to rely on model predictions.
Collecting volunteer annotations to maximise model performance ultimately makes volunteers even more vital because we gain a multiplicative benefit from each annotation; a volunteer annotating one galaxy helps improve a model that annotates all galaxies.

\subsection{\label{sec:finetuning} Foundation models and finetuning}

The models used here are the end result of a research project developing adaptable models for galaxy morphology.
We briefly summarise the computer science motivation and previous progress below.

Transfer learning is the practice of training on one task to do better at a second task.
Image features learned on the first task are hoped to `transfer' (be relevant) to the second task \citep{Lu2015}.
Transfer learning is especially useful where data for the second task is scarce.
This was recognised early on as a useful technique in astronomy \citep{Ackermann2018a,DominguezSanchez2019,Tang2019}.

Separately, models trained simultaneously on a diverse set of related tasks often outperform models trained on any single task \citep{Caruana1997}.
One explanation is that labels for one task can help models learn general image features relevant to other tasks.
Earlier work in this project trained models on multiple morphology tasks in a single survey \citep{Walmsley2022decals} and then expanded to training models on several closely related surveys \citep{walmsleyGalaxyZooDESI2023}.

Foundation models \citep{2021arXiv210807258B,2023arXiv230407193O} combine both transfer learning and multi-task learning. Foundation models involve two model-building phases: `pre-training' on multiple tasks and then `downstream finetuning' where the trained model is adapted to a new task. The hope is that the foundation model learns to extract generally useful image features (from multi-task learning) which are then applied to solve the new task (as in transfer learning). \citet{Walmsley2022practical} found that the multi-survey pretrained model extracted features that were useful for similarity search (finding similar galaxies to a query galaxy), personalised anomaly recommendation (finding galaxies interesting to a specific user), and new morphology tasks. This motivated the release of \texttt{Zoobot} \citep{Walmsley2023zoobot}, the first galaxy foundation models designed to be adapted by other people to new galaxy image tasks. \texttt{Zoobot} is part of a recent trend towards foundation models in astronomy \citep{rozanskiSpectralFoundationModel2023,Leung2023,koblischkeSpectraFMTuningStellar2024,parker_astroclip_2024}. In related work, \cite{Q1-SP049} experiments with applying the foundation model of \cite{smithAstroPTScalingLarge2024} to Q1.

Model `scaling laws' (not to be confused with galaxy scaling laws) describe how model performance predictably increases when increasing any of one variable of data, training compute\footnote{The number of calculations required to train the model, typically measured in floating point operations (FLOPs)}, or parameters, provided the other two variables are plentiful. This appears to be true largely independently of model architecture \citep{Kaplan2020,hoffmannTrainingComputeOptimalLarge2022}. Because foundation models are pretrained on diverse tasks with cumulatively plentiful data, they can take advantage of scaling laws by increasing in parameter size and training compute. This underlies the recent success of `large' language models and recent demand for AI training hardware. \cite{walmsleyScalingLawsGalaxy2024} investigated model scaling laws for galaxy images (see also \citealt{smithAstroPTScalingLarge2024}) and released new `\texttt{Zoobot 2.0}' models trained on $10^8$ volunteer annotations. We use these models here.

The base models used in this work and deployed in the \Euclid pipeline \emph{were not trained on \Euclid data}. They are the \texttt{Zoobot} foundation models introduced in \cite{walmsleyScalingLawsGalaxy2024} and designed to adapt to new tasks and new surveys. They were previously successfully tested on Euclidised HST images morphology in \citet{euclidcollaborationEuclidPreparationXLIII2024}. 
We use the volunteer annotations to learn a linear mapping, equivalent to logistic regression, projecting the image features extracted by the base model onto \Euclid morphology measurements. In neural network terminology, we add a new `head' layer with one unit per morphology answer and freeze the base layers. 

\section{\label{sec:results} Results}

The most intuitive way to demonstrate the quality of visual morphology measurements is visually.

\begin{figure}
    \centering
    \includegraphics[width=\linewidth]{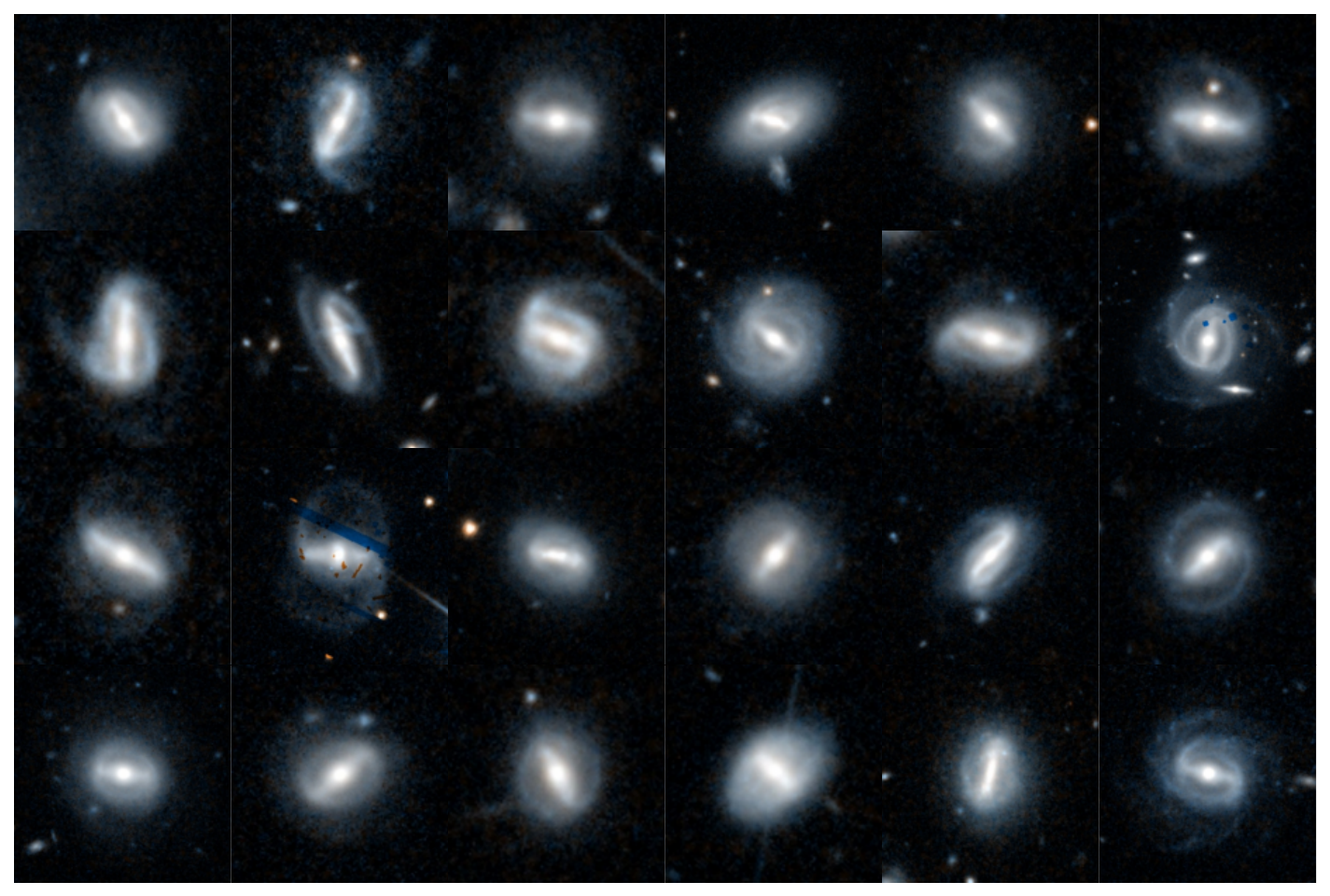}
    \caption{Q1 galaxies predicted as most likely to host strong bars.}
    \label{fig:bar_fraction}
\end{figure}

\begin{figure}
    \centering
    \includegraphics[width=\linewidth]{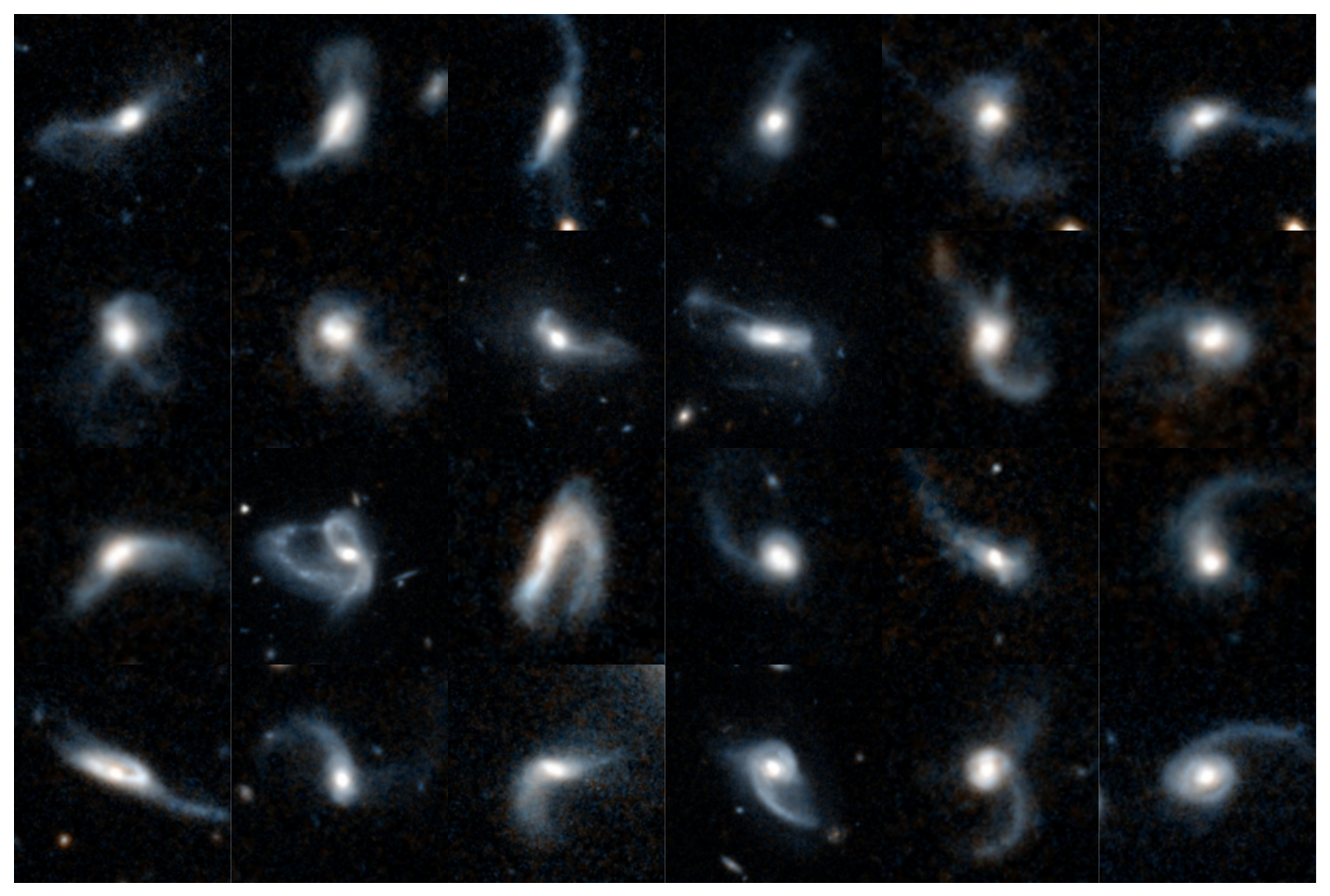}
    \caption{Q1 galaxies predicted as most likely to have a major disturbance (typically interpreted as tidal tails or similar structures, and distinct from ongoing mergers such as close pairs).}
    \label{fig:major_disturbance_fraction}
\end{figure}

\begin{figure}
    \centering
    \includegraphics[width=\linewidth]{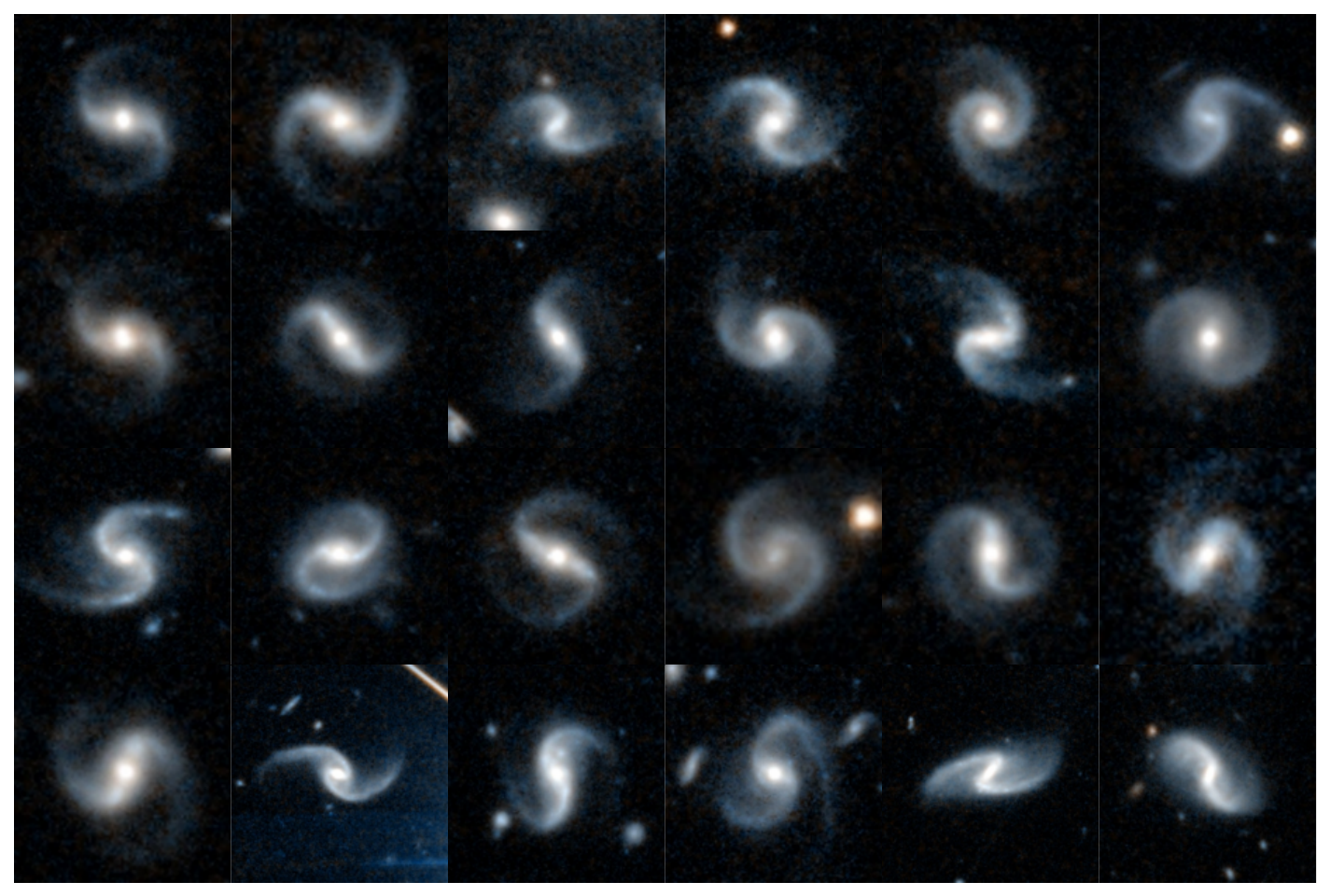}
    \caption{Q1 galaxies predicted as most likely to have exactly two spiral arms.}
    \label{fig:two_arm_fraction}
\end{figure}

\begin{figure}
    \centering
    \includegraphics[width=\linewidth]{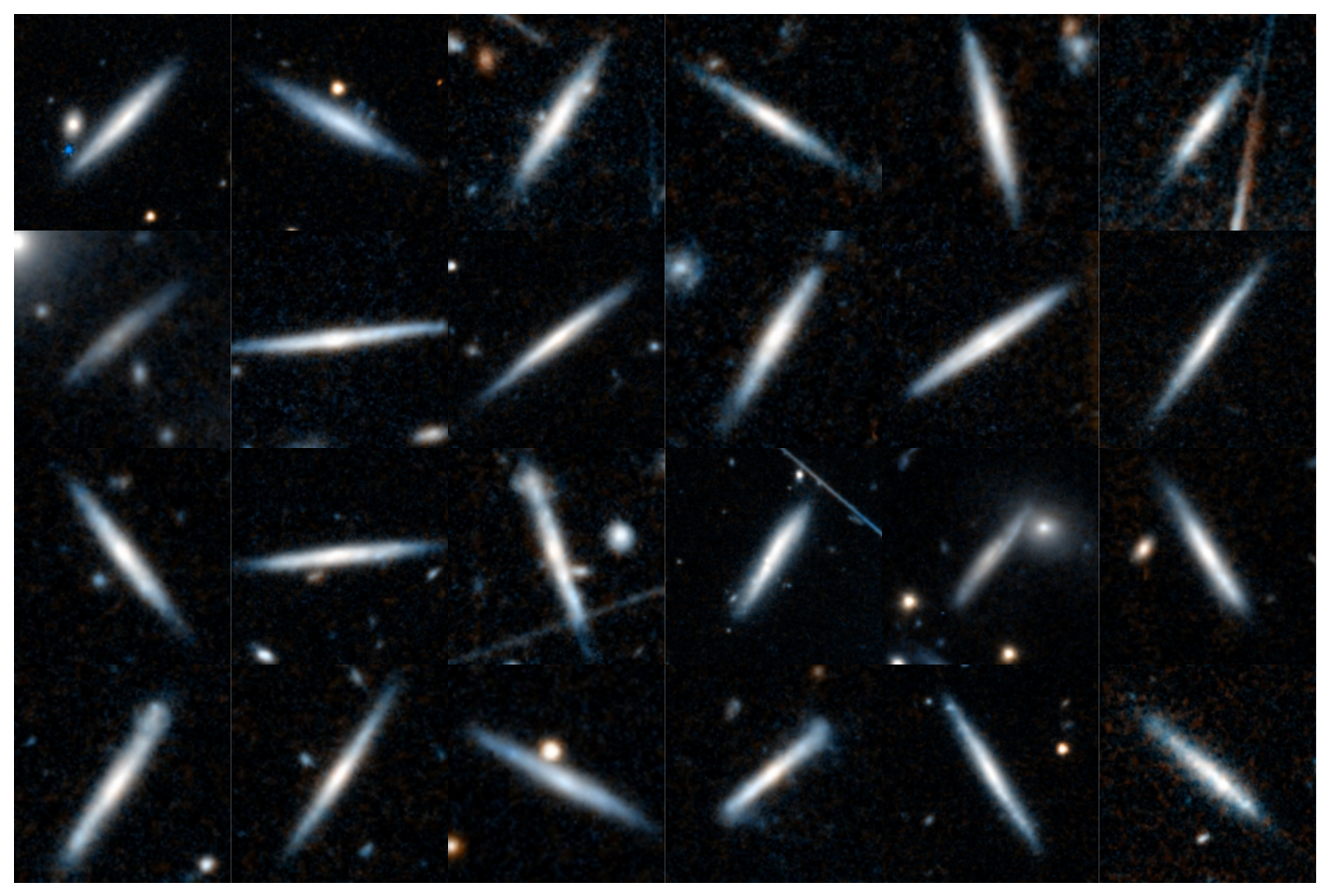}
    \caption{Q1 galaxies predicted as most likely to be bulgeless edge-on disk galaxies. These are likely to be free of recent mergers and hence  are useful laboratories for investigating galaxy and supermassive black hole growth \citep{Simmons2013,smethurstEvidenceNonmergerCoevolution2024}.}
    \label{fig:bulgeless_fraction}
\end{figure}

Figures \ref{fig:bar_fraction},  \ref{fig:major_disturbance_fraction}, and \ref{fig:two_arm_fraction} demonstrate three challenging visual morphology tasks: identifying strong bars, tidal tails, and galaxies with exactly two spiral arms. 
Traditional methods for identifying these features are typically only applied to (relatively) small samples of hundreds to thousands of galaxies, e.g., \cite{Hoyle2011,2017A&A...601A.132G,consolandiAutomatedBarDetection2016,leeBarClassificationBased2020,smithGrandDesignVs2024}.
Figure \ref{fig:bulgeless_fraction} demonstrates identifying bulgeless edge-on disk galaxies. These are of particular scientific interest as they are likely to be free of recent mergers and hence are useful laboratories for investigating galaxy and supermassive black hole growth \citep{Simmons2013,smethurstEvidenceNonmergerCoevolution2024}.

\begin{figure}
    \centering
    \includegraphics[width=\linewidth]{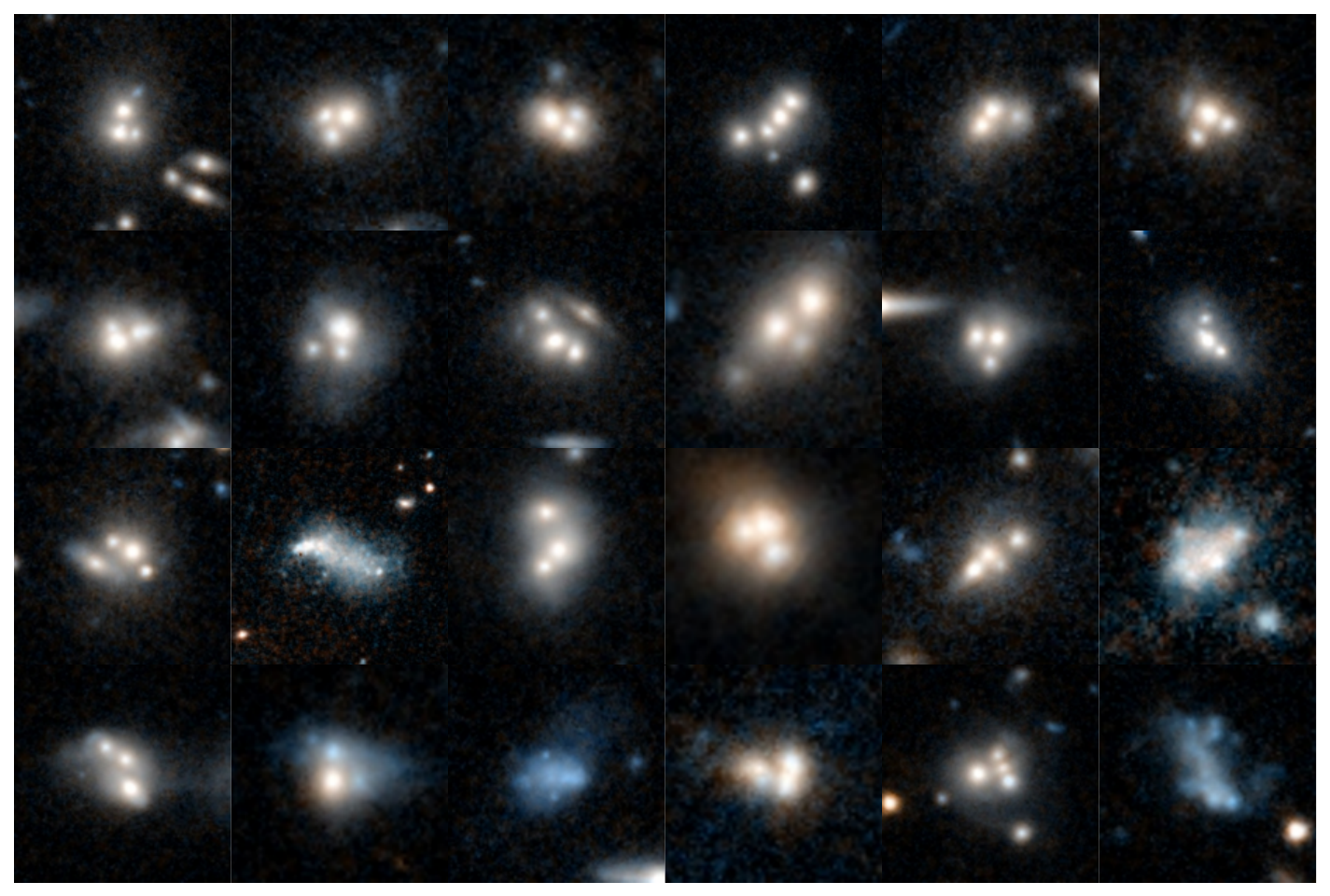}
    \caption{Q1 galaxies predicted as featured but \textit{least} likely to have spiral arms. Volunteers were never asked to annotate clusters or multi-mergers, but \texttt{Zoobot} has nonetheless learned to extract features that identify such galaxies.}
    \label{fig:least_spiral}
\end{figure}

\begin{figure}
    \centering
    \includegraphics[width=\linewidth]{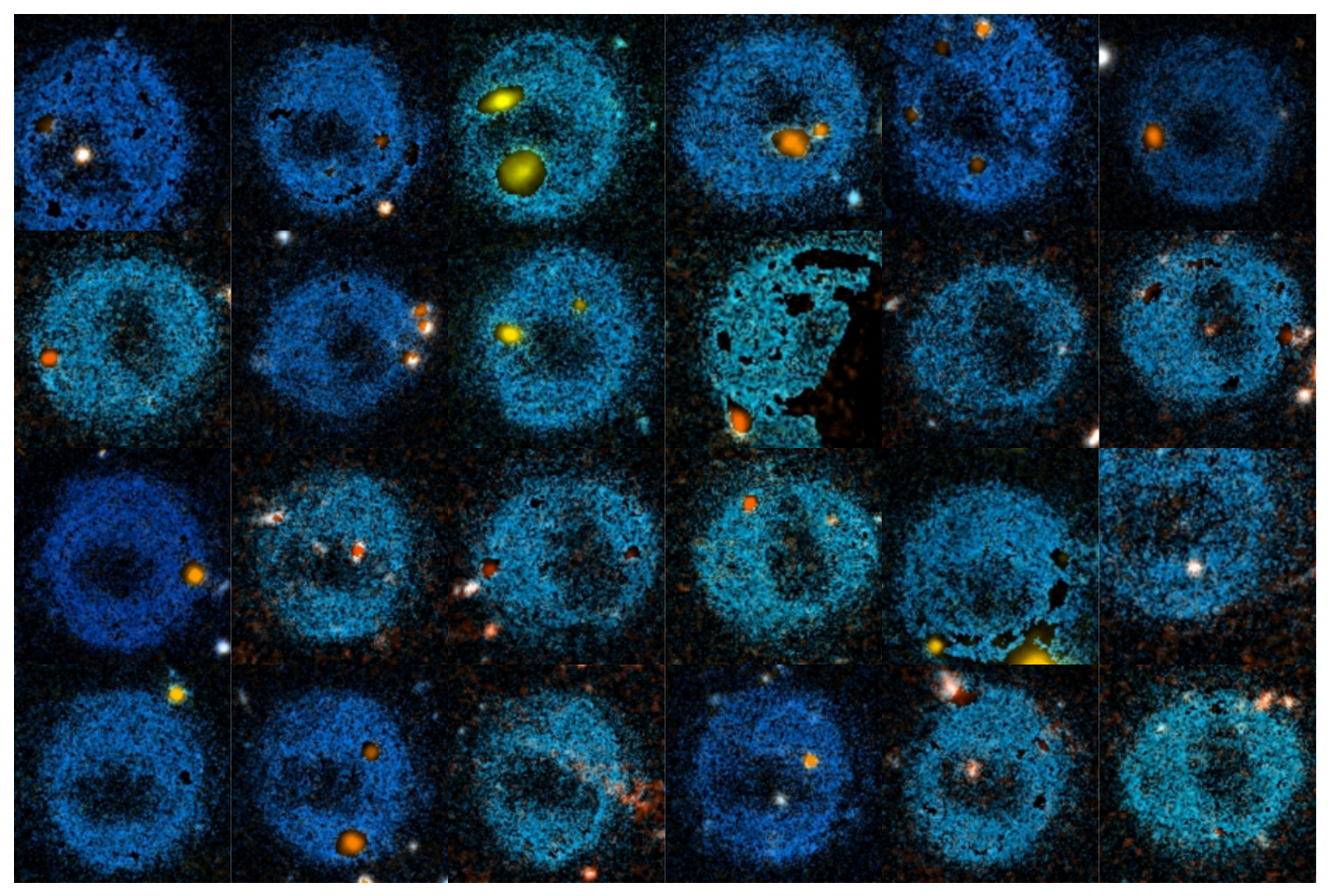}
    \caption{Q1 sources predicted as most likely to be dichrotic ghosts, a well-known artifact caused by internal reflections in telescopes \citep{EuclidSkyVIS,EuclidSkyNISP}.}
    \label{fig:ghost_fraction}
\end{figure}

Our catalogue is also useful for measuring less conventional morphology.
Figure \ref{fig:least_spiral} inverts the previous search for two-armed spirals (Fig. \ref{fig:two_arm_fraction}) and shows the galaxies which are featured but least likely to be two-armed spirals. This identifies galaxies involved in multiple ongoing mergers. This illustrates how \texttt{Zoobot}'s features have generalised beyond the volunteer labels originally used for training; volunteers were not asked to separately identify multiple mergers. Finally, Fig. \ref{fig:ghost_fraction} shows images of dichrotic ghosts, a common artifact \citep{EuclidSkyNISP}. We identify eight categories of problematic images including stars, saturation features, and bright diffraction spikes.

For a quantitative assessment of the performance of our models, we assess their agreement with volunteers on an intensively-annotated subset of intensively-labelled galaxies created for this purpose (\num{3500} galaxies each with 25 annotations).

\begin{figure}
    \centering
    \includegraphics[width=0.9\linewidth]{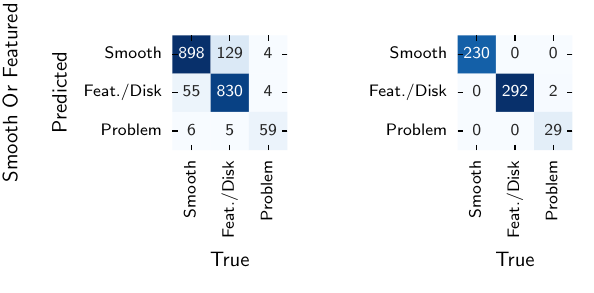}
    \includegraphics[width=0.9\linewidth]{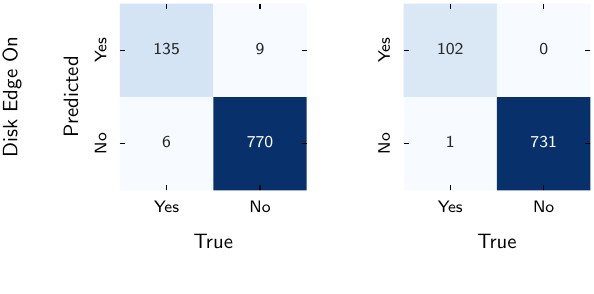}
        \includegraphics[width=0.9\linewidth]{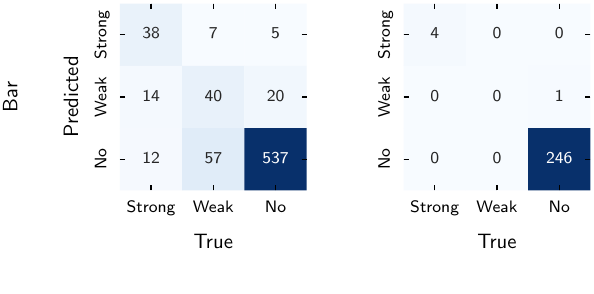}
    \includegraphics[width=0.9\linewidth]{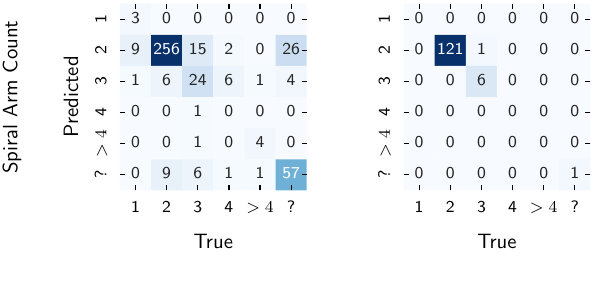}
    \includegraphics[width=0.9\linewidth]{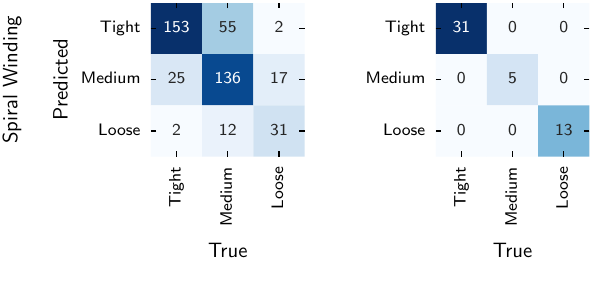}
    \caption{Confusion matrices for the most commonly-used morphology features. Right column includes only galaxies with high confidence (volunteer vote fraction $ > 0.8$) labels. Our model is near-perfect at all questions when evaluated on high-confidence labels. Performance including lower-confidence labels is more mixed, which likely reflects more challenging images for both volunteers and models.}
    \label{fig:cm_smooth}
\end{figure}

We first report classification metrics. These are created by binning the fraction of volunteers giving each answer (for example, if 60\% of volunteers answered `Featured', we bin this label to `Featured').
We report metrics on, both all galaxies and (following \citealt{Sanchez2018}) on galaxies for whom the class label is confidently known (defined as a volunteer vote fraction above 80\% for that answer). Figure \ref{fig:cm_smooth} shows the resulting confusion matrices.

\begin{figure}
    \centering
    \includegraphics[width=\linewidth]{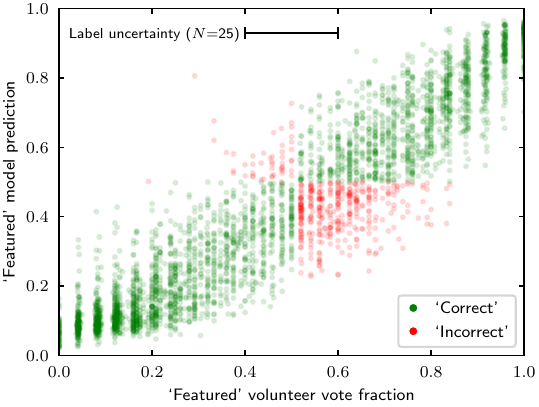}
    \includegraphics[width=\linewidth]{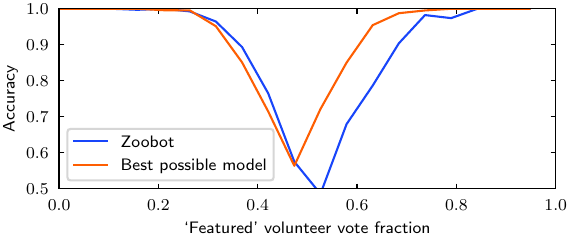}
    \caption{Detailed model performance for the `Smooth or Featured?' question, illustrating the limitations of classification metrics. Above, the measured ($x$-axis) vs. predicted ($y$-axis) `Featured' vote fractions, coloured according to whether the binned prediction equals the binned label (`correct') or not (`incorrect'). Classification errors mostly occur for vote fractions where the true label is uncertain (when the label boundary is within our uncertainty on the vote fraction itself). Below, model accuracy as a function of `featured' vote fraction for either \texttt{Zoobot} or a perfect model (see Sect. \ref{sec:results}), illustrating that \texttt{Zoobot} is near-perfect within the uncertainty on our labels.}
    \label{fig:binning_illustration}
\end{figure}

Our model is near-perfect at all questions when evaluated on high-confidence labels, achieving over 99\% accuracy on 7 of 13 questions and no lower than 95\% accuracy on any question. Performance including lower-confidence labels is more mixed, which likely reflects firstly, more challenging images for both volunteers and models, and secondly, statistical uncertainty in our binned labels. Figure \ref{fig:binning_illustration} illustrates this for the first morphology question (`smooth or featured?'). When the volunteers give a vote fraction decisively skewed to one answer, our model always predicts that answer. As we move to vote fractions near 0.5, the model begins to make nominally incorrect class predictions -- but the binomial uncertainty on the volunteer vote fraction suggests that many binned volunteer labels will fall to one side by chance. 

We quantify this by simulating the predictions of a perfect model. We do this by, for each galaxy, drawing a new set of 25 trials from a binomial distribution with $p$ set to the actual volunteer vote fraction. 
Because the outcome of those new trials includes some uncertainty, the fraction of successful trials is not the same as $p$.
For example, a true vote fraction just below 0.5 (which we should label as 0) will sometimes give a fraction of successful trials above 0.5 and then be given an incorrect label of 1. 
In our analogy, these correspond to galaxies where our perfect model has made the correct prediction of $p$, but where the binomial uncertainty in volunteer responses has caused us to record (by misfortune) the wrong label, and so we incorrectly record the model as wrong.
When compared to this perfect model, 
we find that \texttt{Zoobot} is only 30\% below the best possible classification accuracy (relative error reduction), primarily attributable to a minor bias towards `Featured'.

\begin{figure}
    \centering
    \includegraphics[width=\linewidth]{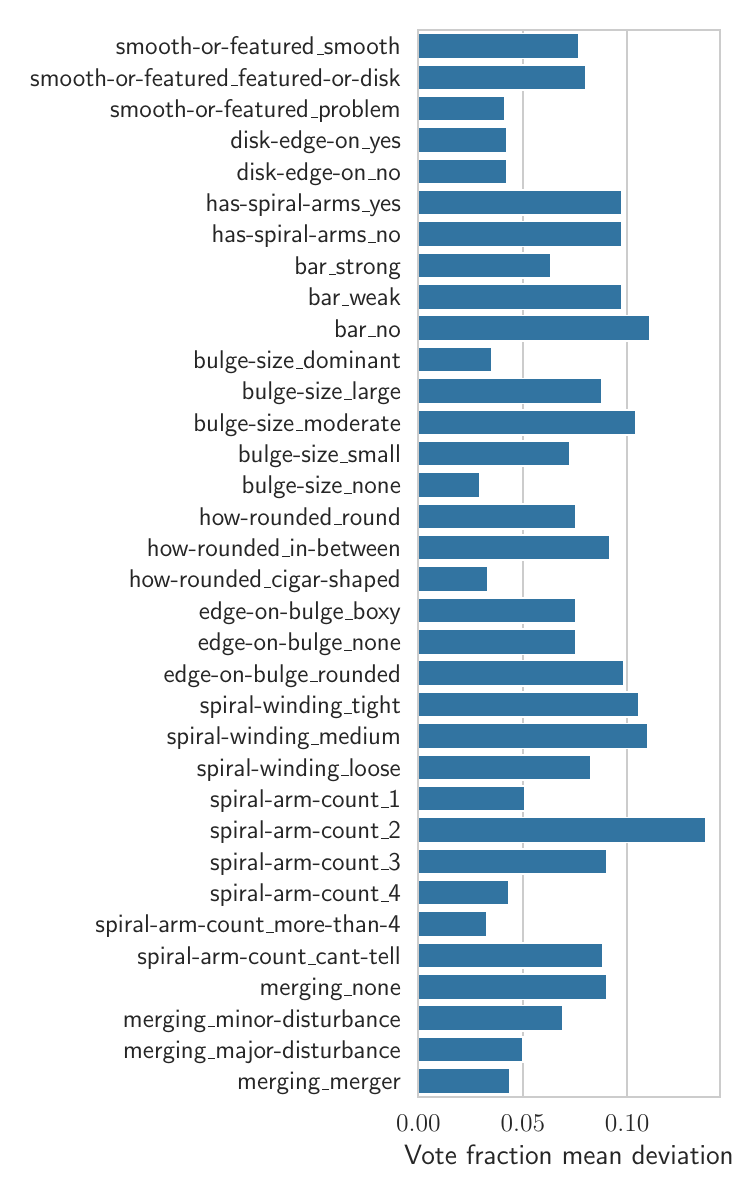}
    \caption{Mean absolute deviation between the predicted and observed volunteer vote fractions. Increasingly detailed questions are increasingly difficult to precisely predict, with the most challenging task being counting spiral arms (17\% error for 2-armed spirals). Errors are typically within 10\%. For conciseness, `clumps' and `problem' questions are not shown.}
    \label{fig:regression metrics}
\end{figure}

We next report regression metrics -- the ability of our models to predict the fraction of volunteers giving each answer. For example, if 60\% of volunteers answering `Featured', our model should predict `0.6'. These avoid the noise introduced by either binning uncertain labels or considering only galaxies with confident labels. Figure \ref{fig:regression metrics} shows the mean absolute deviation between the predicted and observed volunteer vote fractions (excluding the artifact-related questions, for which we have insufficient examples to calculate reliable metrics). \texttt{Zoobot} typically estimates the volunteer vote fraction to within 10\%. Consistent with previous work \citep{walmsleyGalaxyZooDESI2023}, increasingly detailed questions are increasingly difficult to precisely predict, with `edge-on disk' predicted most accurately (4\% error) and `spiral arm count' least accurately (17\% for 2-armed spirals).

\section{\label{sec:data_access} Data access}

All data is available from Zenodo\footnote{\url{https://doi.org/10.5281/zenodo.15002907}}. We also share our data in a machine-learning-friendly format on HuggingFace\footnote{\url{https://huggingface.co/collections/mwalmsley/euclid-67cf5a80e2a93f09e6e4df2c}}. This includes the catalogue, the cutout images, the embeddings (vectors summarising the content of each image), and the models. Documentation is provided at those links; below, we provide a summary.

\subsection{\label{sec:dynamic_catalogue} Catalogues}


The dynamic catalogue contains the columns listed below.
\begin{itemize}
    \item Cross-matching information for the MERge catalogue: object\_id ; tile\_index. 
    \item id\_str. Use to join with embeddings table (below). Formatted like \{release\_name\}\_\{tile\_index\}\_\{object\_id\}. 
    \item Key MERge catalogue columns for convenience: right\_ascension and declination (degrees); kron\_radius; mag\_segmentation (\IE, from flux\_segmentation); segmentation\_area.
    \item Paths to jpg cutouts.
    \item Detailed morphology measurements formatted like 
    \subitem {question}\_{answer}\_fraction\\(e.g. smooth-or-featured\_smooth\_fraction),
    \subitem {question}\_{answer}\_dirichlet\\(e.g. smooth-or-featured\_smooth\_dirichlet).
\end{itemize}

We expect that most catalogue users will be primarily interested in the fraction columns. For example, the column `smooth-or-featured\_smooth\_fraction' includes the fraction of volunteers predicted by \texttt{Zoobot} to give the answer `Smooth' when asked `Is this galaxy smooth or featured?'. They can be combined; for example, `smooth-or-featured\_featured-or-disk\_fraction > 0.5' and `disk-edge-on\_no > 0.5' would select galaxies which are featured and face-on. Catalogue users might use these columns as selection cuts (to investigate galaxies with specific morphologies) or consider how these measurements correlate with other common measurements like mass, star formation rate, location in the cosmic web, etc. There is no `best' choice of cuts, because it depends on your aim; increasing the threshold for any cut will make your sample purer but smaller. We suggest starting generously (with thresholds of 0.5 for most questions, or lower for questions with many or rare answers) and raising your thresholds until the sample reaches your desired purity, as judged from the images.

The pipeline catalogue includes only the Dirichlet values and they are named simply as {question}\_{answer}.
For the Dirichlet columns, each value is a parameter for a Dirichlet distribution. The Dirichlet distribution is the multivariate version of the beta distribution\footnote{If helpful, a visualisation tool for the beta distribution is available at \url{https://homepage.divms.uiowa.edu/~mbognar/applets/beta.html}}.
When both beta parameters (that is, both answers to a morphology question) have low values, the beta distribution is flat, and we are uncertain about the galaxy morphology.
When one answer is high, and one answer is low, we are confident in that high answer.
The `\_dirichlet' columns therefore encode both the predicted vote fraction and the uncertainty on that predicted vote fraction. One can calculate the predicted fraction with $\text{E}(X_i) = \alpha_{i} / \sum{\alpha}$ and the uncertainty with $\text{Var}(X_i) = \alpha_{i} (1-\alpha_{i})/ (1 + \sum{\alpha})$ where $\alpha_{i}$ is the Dirichlet concentration of the chosen answer and $\sum{\alpha}$ is the sum of concentrations for all answers to the chosen question.

The morphology measurements are \texttt{Zoobot} predictions, not volunteer answers. 
\texttt{Zoobot} predicts every answer to every question. To avoid providing measurements where a question is not relevant (for example, answering “how many spiral arms? for a smooth galaxy), we set morphology predictions to NaN where the answer is expected to be not relevant. We define this as a leaf probability (that is, the product of all the vote fractions which led to that question) below 0.5.

The detailed morphology measurements in the pipeline catalogue are released as part of the MER (as in, MERged data, see \citealt{Q1-TP004}) catalogue. The MER catalogue also includes common measurements like photometry, including photometry from other surveys; please refer to the ESA \Euclid Science Archive  website\footnote{\href{https://eas.esac.esa.int/sas/}{https://eas.esac.esa.int/sas/}} for the latest information. 

The detailed morphology measurements in the dynamic catalogue are made outside of the official pipeline. This allows more experimentation and flexibility. For example, the dynamic catalogue can use composite $\IE + \YE$ images, while the official pipeline uses \IE images only. For another example, we can update the model (such as by introducing new labels) and make new predictions at any time. In general, we expect the dynamic catalogue to include the latest `bleeding edge' measurements, while the pipeline catalogue will include slowly-changing measurements that match the release cadence of the \Euclid mission data releases. The difference between the dynamic catalogue and pipeline catalogue should reduce over time as we settle into `normal operations'.

\subsection{Cutouts}

We share cutouts of all galaxies in the catalogue, in two formats. For file-based access, we upload our cutouts as part of our Zenodo archive. For machine learning applications, we also share our cutouts on the HuggingFace Hub along with our embeddings (below). Cutouts are saved in native resolution for storage efficiency; you may wish to resize them to a constant size\footnote{For example, with \texttt{PIL}, you could use \texttt{Image.open(original\_loc).resize((300, 300)).save(new\_loc)}}.

Images for all of the galaxies in our catalogue are created from MERge mosaics, described in \citet{Q1-TP004}
The original Q1 data is available via the ESA \Euclid Science Archive\footnote{\href{https://eas.esac.esa.int/sas/}{https://eas.esac.esa.int/sas/}} and described in \citet{Q1-TP001}.

We also share reference code for creating our images on GitHub\footnote{\url{https://github.com/mwalmsley/bulk-euclid-cutouts}}. This code is primarily intended for making \Euclid cutouts at scale via ESA Datalabs \citep{Datalabscite} but also acts as a public record of our exact process.

\subsection{Embeddings}

We previously (Sect. \ref{sec:finetuning}) described how foundation models aim to extract image features that summarise the visual content of each image. Each feature is an $N$-dimensional coordinate vector (here, $N=512$) locating (embedding) the image in an N-dimensional space. The linear mapping we describe in Sect. \ref{sec:finetuning} aims to learn which volume in this space corresponds to which volunteer answers. But embeddings are also useful for broader tasks such as similarity search, including for galaxy morphology \citep{Stein2021,parker_astroclip_2024,Q1-SP049}. 

The \texttt{Zoobot} embeddings are presented on Zenodo and the HuggingFace Hub as a table with rows of galaxies and columns like ‘feat\_pca\_{n}’, where feat\_pca\_{n} 
is the $N^{\rm th}$ principle component of our higher-dimensional embedding. We include the first 40 components (preserving 94\% variance). We also share the uncompressed higher-dimensional embedding, similarly with columns like ‘feat\_{n}’. Bear in mind that many methods may struggle (either performing poorly due to the `curse of dimensionality' or becoming impractically slow) in high dimensions. 

\begin{figure}
    \centering
    \includegraphics[width=\linewidth]{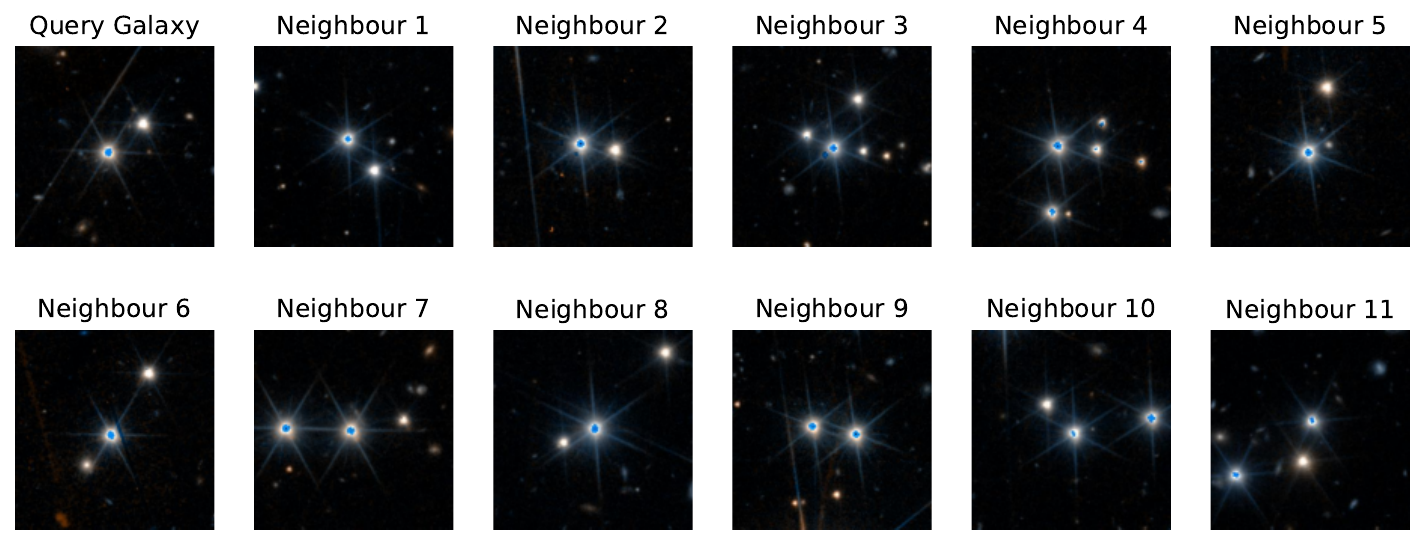}
    \vspace{5mm}
    \vfill
    \includegraphics[width=\linewidth]{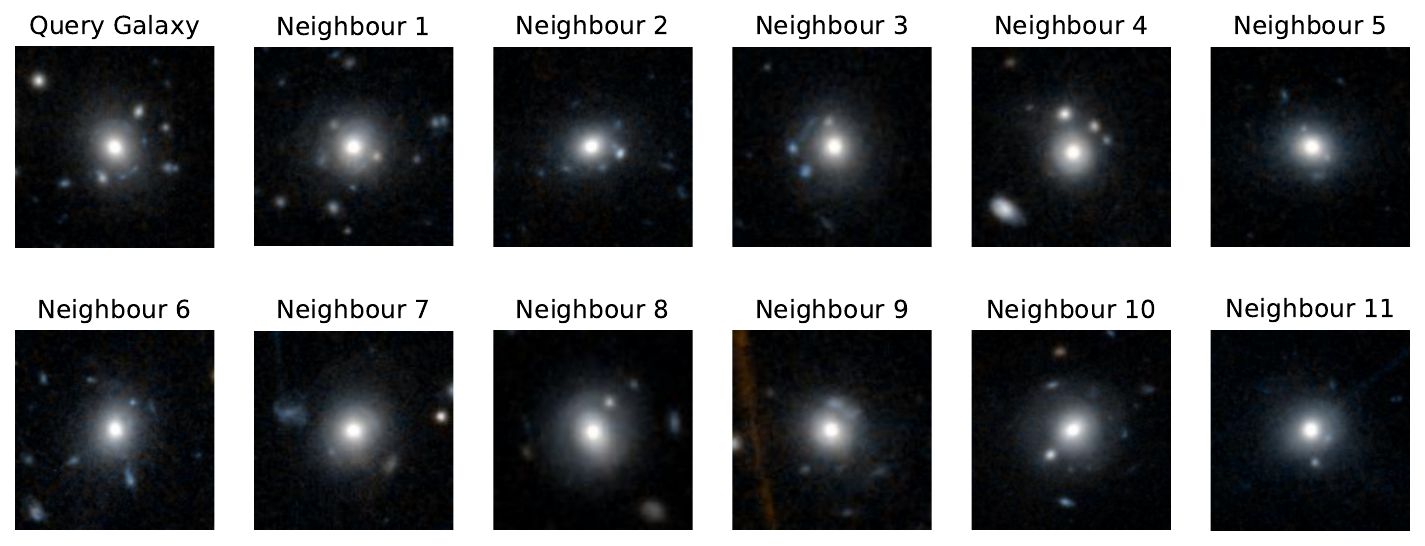}
    \caption{Similarity searches for a pair of stars (above) and strong gravitational lens candidates (below) in Q1.}
    \label{fig:similarity_search}
\end{figure}

Figure \ref{fig:similarity_search} shows two similarity searches made on the \texttt{Zoobot} Q1 embeddings. Note that these embeddings were not created using any \Euclid data (Sect. \ref{sec:finetuning}).

\section{\label{sc:conclusion_and_outlook}  Conclusion and outlook}

Our catalogue provides robust visual morphology measurements (e.g., spiral arm counts, bars, mergers) for the bright and extended galaxies in Q1. These measurements complement traditional morphology measurements \citep{Q1-TP004} and, together, both catalogues provide a comprehensive description of morphology for all Q1 galaxies. Our measurements have already proven useful for addressing a range of science questions; for example, \cite{Q1-SP043} presents new precise measurements of the fraction of barred galaxies out to $z=1$.  We look forward to seeing the work of the wider community.

Deep learning does not automatically resolve the fundamental limitations of visual morphology -- it only makes them scale. We use human responses to images as our ground truth throughout and do not attempt to correct for observational biases such as redshift or angular size\footnote{In analogy to magnitudes, we report apparent and not absolute morphology.}. The catalogue reports what is visible in each image, as a necessary first step to estimating the intrinsic nature of each galaxy. We anticipate that future progress in large-scale galaxy morphology will come not from more accurate models but from models that take a broader interpretation of what learning means (for example, \citealt{cranmerInterpretableMachineLearning2023} and \citealt{wuInsightsGalaxyEvolution2024}) and from the thoughtful combination of models with other essential astronomy tools like simulations and statistics.

Summarising deep learning for survey astronomy, \citet{HuertasCompany2023Dawes} write that `the majority of works are still at the proof-of-concept stage'.
In this work, we show that one month of volunteer effort is sufficient to make a science-ready morphology catalogue for $10^8$ in \Euclid, not just as a proof-of-concept but by creating the first rows of that catalogue. The foundation model used was only trained on \Euclid data for a final linear mapping. We used it to create a galaxy morphology catalogue; we might have chosen another image task. A similar approach has proven successful on strong lenses \citep{Q1-SP048}, active galactic nuclei \citep{Q1-SP015}, star-forming clumps \citep{poppTransferLearningGalaxy2024}, mergers \citep{margalef-bentabolGalaxyMergerChallenge2024}, tidal features \citep{omoriGalaxyMergersSubaru2023,oryanHarnessingHubbleSpace2023}, anomaly searches \citep{lochnerAstronomalyProtegeDiscovery2024}, segmentation \cite{sazonova_structural_2022}, etc. We encourage the reader to experiment with using the tools\footnote{\href{https://github.com/mwalmsley/zoobot}{\texttt{https://github.com/mwalmsley/zoobot}}} behind our catalogue -- the foundation models, and the code to adapt them -- to create exactly the catalogue needed for each science case.

\begin{acknowledgements}

The data in this paper are the result of the efforts of the Galaxy Zoo volunteers, without whom none of this work would be possible. Their efforts are individually acknowledged at \url{http://authors.galaxyzoo.org}. 

 The Dunlap Institute is funded through an endowment established by the David Dunlap family and the University of Toronto. MHC acknowledges support from the State Research Agency (AEIMCINN) of the Spanish Ministry of Science and Innovation under the grants “Galaxy Evolution with Artificial Intelligence" with reference PGC2018-100852-A-I00 and "BASALT" with reference PID2021-126838NBI00.

 This publication uses data generated via the Zooniverse.org platform, development of which is funded by generous support, including a Global Impact Award from Google, and by a grant from the Alfred P. Sloan Foundation.

\AckEC  

\AckQone

\AckDatalabs

\end{acknowledgements}

\bibliography{my, Euclid, Q1}

%

\begin{appendix}
  \onecolumn 

\section{Classification metric tables}

This section records the classification metrics on, split between all galaxies and high-confidence ($p > 0.8$) galaxies.

\begin{table}
\caption{Classification metrics on all galaxies. See Sect. \ref{sec:results}.}
\renewcommand{\arraystretch}{1.5}
\centering
\begin{tabular}{cccccc}
Question & Count & Accuracy & Precision & Recall & F1 \\
\hline \hline
Smooth Or Featured & 3546 & 0.9041 & 0.9041 & 0.9041 & 0.9041 \\
Disk Edge On & 1648 & 0.9806 & 0.9806 & 0.9806 & 0.9806 \\
Has Spiral Arms & 1308 & 0.9029 & 0.9029 & 0.9029 & 0.9029 \\
Bar & 1308 & 0.8471 & 0.8471 & 0.8471 & 0.8471 \\
Bulge Size & 1308 & 0.8242 & 0.8242 & 0.8242 & 0.8242 \\
How Rounded & 1573 & 0.9091 & 0.9091 & 0.9091 & 0.9091 \\
Edge On Bulge & 173 & 0.8902 & 0.8902 & 0.8902 & 0.8902 \\
Spiral Winding & 794 & 0.7292 & 0.7292 & 0.7292 & 0.7292 \\
Spiral Arm Count & 794 & 0.7746 & 0.7746 & 0.7746 & 0.7746 \\
Merging & 3463 & 0.9090 & 0.9090 & 0.9090 & 0.9090 \\
Clumps & 1308 & 0.7638 & 0.7638 & 0.7638 & 0.7638 \\
Problem & 98 & 0.8673 & 0.8673 & 0.8673 & 0.8673 \\
Artifact & 24 & 0.7083 & 0.7083 & 0.7083 & 0.7083 \\
\end{tabular}
\end{table}

\begin{table}
\caption{Classification metrics on high-confidence ($p > 0.8$) galaxies. See Sect. \ref{sec:results}.}
\renewcommand{\arraystretch}{1.5}
\centering
\begin{tabular}{cccccc}
Question & Count & Accuracy & Precision & Recall & F1 \\
\hline \hline
Smooth Or Featured & 1136 & 0.9974 & 0.9974 & 0.9974 & 0.9974 \\
Disk Edge On & 1478 & 0.9993 & 0.9993 & 0.9993 & 0.9993 \\
Has Spiral Arms & 787 & 0.9848 & 0.9848 & 0.9848 & 0.9848 \\
Bar & 453 & 0.9934 & 0.9934 & 0.9934 & 0.9934 \\
Bulge Size & 120 & 1.0000 & 1.0000 & 1.0000 & 1.0000 \\
How Rounded & 875 & 0.9977 & 0.9977 & 0.9977 & 0.9977 \\
Edge On Bulge & 69 & 1.0000 & 1.0000 & 1.0000 & 1.0000 \\
Spiral Winding & 85 & 0.9882 & 0.9882 & 0.9882 & 0.9882 \\
Spiral Arm Count & 235 & 0.9830 & 0.9830 & 0.9830 & 0.9830 \\
Merging & 1090 & 1.0000 & 1.0000 & 1.0000 & 1.0000 \\
Clumps & 419 & 0.9523 & 0.9523 & 0.9523 & 0.9523 \\
Problem & 40 & 0.9750 & 0.9750 & 0.9750 & 0.9750 \\
Artifact & 4 & 1.0000 & 1.0000 & 1.0000 & 1.0000 \\
\end{tabular}
\end{table}

\label{LastPage}

\end{appendix}

\end{document}